\documentclass[aps,prd,superscriptaddress,nofootinbib,eqsecnum,twocolumn]{revtex4-1}

\pdfoutput=1

\usepackage{graphicx}
\usepackage{bm}
\usepackage{dcolumn}
\usepackage{amssymb} 
\usepackage{float}
\usepackage{amsmath} 
\usepackage{placeins}
\usepackage{epsfig} 
\usepackage{subfigure}
\usepackage[english]{babel}
\usepackage[latin1]{inputenc}
\usepackage[T1]{fontenc} 
\usepackage{eucal}
\usepackage{hyperref}
\usepackage{verbatim}
\usepackage{latexsym}
\usepackage{color,xcolor}
\usepackage{ulem}
\usepackage{soul} 

\DeclareFontFamily{OT1}{pzc}{}
\DeclareFontShape{OT1}{pzc}{m}{it}{<-> s * [1.1] pzcmi7t}{}

\newcommand{\termchancery}{\fontfamily{pzc}\selectfont}

\hypersetup{colorlinks=true,linkcolor=orange,filecolor=magenta,urlcolor=cyan,citecolor=cyan}

\begin{document}

\title{Medium separation scheme effects on the magnetized and cold two-flavor superconducting quark matter} 
\author{Francisco X. Azeredo}  
\email[Electronic address: ]{francisco.azeredo@acad.ufsm.br}
\affiliation{Departamento de F\'{\i}sica, Universidade Federal de Santa Maria, 97105-900, Santa Maria, Rio Grande do Sul, Brazil}

\author{Dyana C. Duarte} 
\email[Electronic address: ]{dyana.duarte@ufsm.br }
\affiliation{Departamento de F\'{\i}sica, Universidade Federal de Santa Maria, 97105-900, Santa Maria, Rio Grande do Sul, Brazil}

\author{Ricardo L. S. Farias}  
\email[Electronic address: ]{ricardo.farias@ufsm.br}
\affiliation{Departamento de F\'{\i}sica, Universidade Federal de Santa Maria, 97105-900, Santa Maria, Rio Grande do Sul, Brazil}
\affiliation{Center for Nuclear Research, Department of Physics, Kent State University, Kent, Ohio 44242 USA}

\date{\today}

\begin{abstract}
We analyze the impact of the medium separation scheme (MSS) on two-flavor color superconducting (2SC) dense quark matter under the influence of a constant external magnetic field. The effects of the proper treatment of the model divergences are examined through a comparison of different approaches, including the combined implementation of the magnetic field independent regularization (MFIR) and the MSS, as well as the standard use of smooth form factors. Our findings for the Nambu-Lasinio model emphasize the critical role of properly separating medium effects from vacuum contributions in the model. The combined MFIR-MSS scheme suppresses spurious unphysical oscillations, often misinterpreted in the literature as de Haas-van Alphen oscillations, and ensures the correct high-density behavior of the diquark condensate. Furthermore, within the MSS framework, the magnetization remains positive across the explored parameter space, in sharp contrast with the behavior obtained in the traditional approach.

\end{abstract}

\maketitle


\section{Introduction}

The possible existence of deconfined quark matter in the core of neutron stars~\cite{Annala:2019puf} has shed new light on the understanding of strongly interacting matter described by QCD at high densities. The expectation that quark matter becomes color superconducting at very high densities has been thoroughly reviewed in the literature~\cite{Bailin:1983bm}. Perturbative calculations were later confirmed within a controlled weak-coupling framework in Ref.~\cite{PhysRevD.59.094019}, where the authors derived the parametric form of the superconducting gap.
Nevertheless, these perturbative approaches are not applicable to the intermediate density regime, which is relevant for the phenomenology of compact objects.
Therefore, alternative approaches must be developed to explore the properties of matter in this regime. From a theoretical perspective, much of our current understanding of the QCD phase structure largely stems from first-principles Monte Carlo simulations on the lattice~\cite{Laermann:2003cv,Parotto:2023mhh}, although the well-known sign problem limits the applicability of this approach at finite chemical potential (see~\cite{Nagata:2021ugx} for a recent review). Consequently, our understanding of the behavior of cold and dense matter in the intermediate density regime relies heavily on the development of effective models. Chiral models such as the Nambu--Jona-Lasinio (NJL)~\cite{Klevansky:1992qe, Buballa:2003qv} and the linear sigma model~\cite{Ayala:2014jla,Andersen:2024qus} aim to capture the essential features of strongly interacting matter by incorporating key symmetries and physical constraints of QCD, providing valuable insights into the equation of state and phase transitions relevant to compact stars and other dense systems.

In Refs.~\cite{ALFORD1998247,PhysRevLett.81.53}, it was shown that fermion spectrum gaps are around 100 MeV, potentially extending color superconducting phases to finite temperatures in the QCD phase diagram, with a rich phase structure depending on the temperature and density scales~\cite{Duarte:2018kfd,ALFORD1999443,PhysRevLett.91.242301}. At moderate baryon densities, the system is dominated by the two-flavor superconducting (2SC) phase, which is the focus of this work. Astrophysical analyses have recently constrained the equation of state of dense quark matter using neutron star observations, effective field theory, and perturbative QCD, thereby placing upper bounds on the superconducting gap~\cite{Kurkela:2024xfh}. Comprehensive reviews can be found in~\cite{Rajagopal:2000wf,Rischke:2003mt,Alford:2001dt}, and first-principles studies of color superconductivity at asymptotic densities in~\cite{PhysRevD.59.094019,SHOVKOVY1999189}. Color superconducting phases are relevant for compact stars, which are expected to reach core densities of $\sim 10$ times the nuclear density $\rho_0$ ($\rho_0 \approx 0.16$ fm$^{-3}$), where such phases might emerge. The absorption of electron and muon neutrinos in the 2SC phase has been investigated in a recent study using a self-consistent NJL model treatment, revealing that in equilibrated 2SC matter the neutrinos can form a degenerate gas with a mean free path of only a few meters, relevant for neutron star mergers~\cite{Alford:2025jtm}.

Color superconducting phases are particularly relevant in astrophysical contexts such as magnetars, where dense matter is subjected to strong magnetic fields. In these extreme environments, the standard electromagnetic Meissner effect may be absent for certain nonspherical superconducting phases, thereby allowing the magnetic flux to penetrate without significant energy cost~\cite{Feng:2009vt}. Earlier investigations of magnetized color-superconducting quark matter have demonstrated that strong magnetic fields can significantly influence both pairing dynamics and the phase structure, modifying the behavior of the gap and the equation of state~\cite{Abhishek:2018xml}. Within the two-flavor sector, rotated charges and anisotropic superconductivity have been investigated in Refs.~\cite{Coppola:2017edn,Cao:2015xja}, emphasizing the necessity of incorporating electric and color neutrality constraints for more realistic descriptions.
Furthermore, intense magnetic fields have been shown to induce oscillatory behavior in the gap parameters~\cite{Fukushima:2007fc,Noronha:2007wg}, emphasizing the nontrivial interplay between magnetic fields and superconductivity. Beyond gap oscillations, more exotic phenomena may arise, such as the formation of inhomogeneous gluon condensates, which mitigate chromomagnetic instabilities in neutral two-flavor superconductors, and the emergence of rotated magnetic fields~\cite{Ferrer:2006ie,Ferrer:2007uw,Yuan:2024ajk}. These effects are crucial for accurately modeling compact stars, especially magnetars. In the three-flavor case, intense magnetic fields may induce new phases, such as the magnetic color-flavor-locked (MCFL) and Paramagnetic-Color-Flavor-Locked (PCFL) phases, wherein both the pairing structure and the spectrum of collective excitations are significantly modified~\cite{Ferrer:2007iw,Feng:2011fj,Paulucci:2010uj}. The alignment of Cooper pair magnetic moments in these phases can lead to the formation of additional condensates, thereby enhancing the pairing energy in the presence of strong fields. From a theoretical perspective, renormalization-group analyses have elucidated the role of long-range magnetic interactions in color superconductivity at asymptotically high densities, leading to a non-BCS scaling of the superconducting gap, $\Delta \sim \mu\, g^{-5} \exp(-c/g)$~\cite{PhysRevD.59.094019,Hsu:1999mp}, where $c$ is a constant. These insights are essential for the construction of consistent models of dense QCD matter under extreme magnetic conditions. Moreover, it has been proposed that strong background fields could induce an electromagnetic superconducting state of the vacuum through the condensation of charged vector $\rho$ mesons, suggesting an analogy between QCD and condensed matter systems~\cite{Chernodub:2010qx,Chernodub:2012tf}. However, recent calculations based on extended NJL models~\cite{Carlomagno:2022inu,Carlomagno:2022arc}, which consistently include Schwinger phases and proper wave functions for $B\neq 0$, have found no evidence of such condensation in the range of magnetic fields considered, thus disfavoring the occurrence of this mechanism.

In this work, we employ the NJL model to study diquark condensation in cold and dense quark matter under an external magnetic field. Given the nonrenormalizable nature of the NJL model, the regularization procedure must be handled with particular care. In this class of effective models, UV divergent momentum integrals must be handled, typically by introducing a sharp cutoff $\Lambda$, which is regarded as a model parameter fitted to physical observables. This cutoff defines an energy scale beyond which the model cannot be trusted. Standard procedures often mix vacuum and medium effects, introducing dependencies on chemical potential, temperature, and external fields. Given that the NJL model serves as an effective approximation of QCD, it is reasonable to impose that regularization affects only vacuum-dependent terms. This principle was first applied in studies of color superconductivity~\cite{Farias:2005cr} and has since been employed in different contexts to describe the QCD phase diagram. The separation, termed the {\it medium separation scheme} (MSS), contrasts with the traditional regularization scheme (TRS), which applies the cutoff to all divergent integrals that include physical quantities that depend on medium effects. One of the main motivations for this study is the evidence that the diquark condensate $\Delta$ increases with baryon chemical potential $\mu_B$. While lattice QCD simulations for $N_c = 3$ are hindered by the sign problem~\cite{Karsch:2001cy}, studies for $N_c = 2$~\cite{Kogut:2001na,Braguta:2016cpw}, chiral perturbation theory (ChPT)~\cite{Kogut:2000ek} and the renormalization-group consistent treatment applied to the NJL model~\cite{Gholami:2024diy} show this increasing behavior of $\Delta$ with $\mu_B$. Using TRS, the diquark condensate starts to decrease at a value of chemical potential smaller than the energy scale $\Lambda$, and eventually vanishes. In contrast, employing MSS, a consistently increasing condensate is predicted, in agreement with other approaches.

This approach has been shown to reconcile NJL model predictions with lattice QCD results for the chiral transition in the presence of a chiral imbalance~\cite{Farias:2016let}. It has also been applied to NJL models with diquark interactions, showing significant qualitative and quantitative differences in the phase structure of cold and dense nuclear matter compared to traditional regularization methods~\cite{Duarte:2018kfd}. Furthermore, MSS has been employed in studies of the QCD equation of state at zero temperature and finite isospin density, improving agreement with lattice QCD and chiral perturbation theory~\cite{Avancini:2019ego}. Extending this analysis to the three-flavor NJL model, MSS has shown good agreement with lattice simulations for low isospin densities, although at higher densities, the two-flavor NJL model unexpectedly provides a better fit~\cite{Lopes:2021tro}. More recently, it has been shown that within the MSS the speed of sound peak in isospin-asymmetric and two-color QCD superfluid emerges as a natural prediction of the NJL model, in excellent agreement with state-of-the-art lattice data~\cite{Lopes:2025rvn, Pasqualotto:2025kpo}. Additionally, the scheme has been incorporated into the Polyakov-NJL model, where a new parametrization of the Polyakov-loop potential, dependent on temperature and chiral chemical potential, yields results consistent with lattice simulations and analytical results from ChPT~\cite{Azeredo:2024sqc}. Moreover, the combination of MSS with the Optimized Perturbation Theory (OPT) method~\cite{Okopinska:1987hp,Duncan:1988hw} has been recently explored beyond the large-$N_c$ limit, revealing that the MSS scheme is essential to recover the correct behavior of the pseudocritical temperature under chiral imbalance and to ensure consistency with lattice QCD predictions~\cite{daSilva:2025koa}. In a related context, the role of the MSS in the violation of the conformal limit within effective descriptions of dense QCD matter at finite isospin density, two-color QCD, and two-flavor color-superconducting quark matter, has also been investigated~\cite {sym18020220}.

Over the recent past, significant attention has been devoted to the generation of magnetic fields in noncentral heavy-ion collisions, where the field strength can reach up to $10^{20}$ G, and which can significantly influence the QCD phases~\cite{Bzdak:2012fr,McLerran:2013hla}. The study of QCD under strong magnetic fields has expanded across different contexts, including magnetars, neutron star mergers, and early Universe physics. For comprehensive overviews of the phase structure of QCD in strong magnetic backgrounds, encompassing effective model approaches, lattice QCD results, and different physical contexts in which magnetic fields play a relevant role, see Refs.~\cite{Andersen:2014xxa,Adhikari:2024bfa,Endrodi:2024cqn,Miransky:2015ava,Bandyopadhyay:2020zte} and references therein. In this context, accurately modeling the behavior of quark matter under strong magnetic fields requires a careful choice of regularization schemes to avoid unphysical artifacts and ensure reliable predictions, especially when working with nonrenormalizable models. The choice of an appropriate regularization scheme is essential for accurately describing magnetized quark matter, as highlighted in previous studies~\cite{PhysRevD.99.116002,Allen:2015paa,Duarte:2015ppa,Duarte:2017nzv}. As an example, one may cite a recent work~\cite{Endrodi:2019whh}, where the authors fixed the running coupling of the two-flavor Polyakov extended NJL model (PNJL) by matching its predictions with LQCD results for the baryon spectrum in a strong magnetic background. This approach, which preserves the expected decrease of the pseudocritical temperature, $T_{pc}$, with the increase of the magnetic field for both chiral and deconfinement transitions, relies on Schwinger's proper time formalism~\cite{PhysRev.82.664}, where all integrals are regularized without separating vacuum and thermomagnetic contributions. However, while this method ensures consistency with LQCD predictions at moderate temperatures, it can lead to inconsistencies at higher temperatures or finite baryon chemical potentials, such as unphysical oscillations in magnetization and imaginary meson masses~\cite{PhysRevD.81.016007, PhysRevD.86.085042, Ruivo:2010ff,Costa:2010zw}. Different regularization schemes have been proposed to address these issues, with the most prominent being the {\it magnetic field independent regularization} (MFIR), proposed in~\cite{PhysRevD.61.025005,Ebert:2003yk} and applied in~\cite{PhysRevC.79.035807}, which isolates vacuum divergences from medium contributions and removes unphysical oscillations~\cite{Ebert:2003yk,PhysRevD.61.025005}. Additionally, variations in handling the thermal contribution, such as the standard proper time and the thermomagnetic regulated proper time schemes~\cite{Endrodi:2019whh}, affect the convergence to the Stefan-Boltzmann limit, the behavior of thermal effective quark masses, and can reach possible regimes with causality violation \cite{PhysRevD.107.096017}.

Given these challenges, this work aims to analyze different regularization approaches and identify those that yield physically reliable results in nonrenormalizable theories. Furthermore, it is crucial that in this work we are applying, for the first time in the context of color superconductivity within the NJL model, the combination of MFIR and MSS to ensure a consistent treatment of both magnetic field and dense matter effects. This approach is crucial for obtaining physically reliable predictions, as will be detailed later.
 This paper is organized as follows. In Sec.~\ref{Sec:model}, we introduce the NJL model at finite density and under the influence of an external magnetic field. Special attention is dedicated to the regularization schemes that separate the medium and magnetic field contributions from divergent integrals. In Sec.~\ref{Sec:results}, we analyze the application of these regularization schemes, emphasizing their importance in reproducing the expected behavior of physical quantities and comparing results with previous findings in the literature. Finally, in Sec.~\ref{Sec:remarks}, we summarize our findings and present the concluding remarks.

\section{The NJL model and regularization}
\label{Sec:model}

In this work, we consider the SU(2)$_f$ NJL model including scalar-pseudoscalar and color pairing interactions. The Lagrangian density in the presence of an external magnetic field can be written as 
\begin{eqnarray}
\mathcal{L} & = & \bar{\psi}\left[i\gamma^{\mu}\left(\partial_{\mu}-i\tilde{e}\tilde{\mathcal{Q}}\tilde{A}_{\mu}\right)+\hat{\mu}\gamma^{0}-\hat{m}\right]\psi
\nonumber\\
&+&G_{S}\left[\left(\bar{\psi}\psi\right)^{2}+\left(\bar{\psi}i\gamma_{5}\vec{\tau}\psi\right)^{2}\right] 
\nonumber\\
 & + & G_{D}\left[\left(i\bar{\psi}^{C}\epsilon_{f}\epsilon_{c}^{3}\gamma_{5}\psi\right)\left(i\bar{\psi}\epsilon_{f}\epsilon_{c}^{3}\gamma_{5}\psi^{C}\right)\right],
\end{eqnarray}
where the quark fields with two flavors $\psi=(u,d)^T$ are charge-conjugate spinors, with $\psi^C = C\bar{\psi}^T$, $\bar{\psi}^C = \psi^T C$ ($C = i\gamma^2\gamma^0$ is the charge conjugation matrix), and $\vec{\tau} = (\tau_1,\tau_2,\tau_3)$ are the Pauli matrices. The current quark mass matrix and chemical potential matrices are defined as
$\hat{m}=\text{diag}\left(m_{u},m_{d}\right)$, and $\hat{\mu}=\left(\mu_{ur},\mu_{ug},\mu_{ub},\mu_{dr},\mu_{dg},\mu_{db}\right)$, respectively. In the isospin-symmetric limit, we assume $m_u = m_d\equiv m_c$ and $\mu_i \equiv \mu$.
Moreover, $\left(\epsilon_{c}^{3}\right)^{ab}=\left(\epsilon_{c}\right)^{3ab}$
and $\left(\epsilon_{f}\right)^{ij}$ are the antisymmetric matrices
in color and flavor spaces, respectively. In this equation, one may see that the coupling of the quarks with the magnetic field $\tilde{\mathcal{A}}_{\mu}$ is implemented through the covariant derivative in terms of the rotated fields. In the presence of a nonvanishing superconducting gap $\Delta$, the photon acquires a finite mass~\cite{Allen:2015paa}. However, as shown in~\cite{Alford:1999pb}, there is a linear combination of the photon and a gluon field that remains massless. The associated rotated charge matrix $\tilde{\mathcal{Q}}$ is given by $\tilde{\mathcal{Q}} = \mathcal{Q}_f\otimes1_c - 1_f\otimes\left(\frac{\lambda_8}{2\sqrt{3}}\right)$, where $\mathcal{Q}_f = \text{diag}(2/3,-1/3)$ and $\lambda_8 = \text{diag}(1,1,-2)/\sqrt{3}$ is the color quark matrix.
Therefore, in the six-dimensional flavor-color representation, the rotated charges for the different quarks are $u_r = u_g = 1/2, u_b = 1, d_r = d_g = -1/2, d_b = 0$. 
Considering a constant and static magnetic
field in the 3-direction, $\tilde{A}_{\mu}=\delta_{\mu2}x_{1}B$, we have a mixture of the electromagnetic field and color fields. The rotated unit charge $\tilde{e} = e\cos(\theta)$, where $\theta$ is the mixing angle, is estimated to be of the order of 1/20~\cite{Gorbar:2000ms}. To simplify notation, we will use $e$ instead of $\tilde{e}$ in what follows. Lastly, $G_S$ and $G_D$ are the scalar-pseudoscalar and diquark coupling constants.

In mean field approximation the corresponding thermodynamic potential, for a given temperature
$T$, reads
\begin{eqnarray}
\Omega_{T} & = & \frac{\left(M-m_{c}\right)^{2}}{4G_{S}}+\frac{\Delta^{2}}{4G_{D}}-\int\frac{d^{3}\vec{p}}{\left(2\pi\right)^{3}}\left[f\left(E_{p,0}^{+}\right)+f\left(E_{p,0}^{-}\right)\right]\nonumber \\
 &  & -\frac{eB}{8\pi^{2}}\sum_{n=0}^{\infty}\alpha_{n}\int\limits_{-\infty}^{+\infty}dp_{z}\bigg[f\left(E_{p,1}^{+}\right)+f\left(E_{p,1}^{-}\right) \nonumber\\
 &&+2f\left(E_{p,\frac{1}{2}}^{+}\right) 
 + 2f\left(E_{p,\frac{1}{2}}^{-}\right)\bigg],
 \label{OmegaT}
\end{eqnarray}
where $f(x) = x+2T\ln\left(1+e^{-x/T}\right)$, $\alpha_n = 2-\delta_{n,0}$ takes into account the degeneracy of Landau levels, and the dispersion relations are 
\begin{align}
    &E^{\pm}_{p,0} = \sqrt{\vec{p}^2 + M^2} \pm \mu\,, \\ 
    &E^{\pm}_{p,1} = \sqrt{p_z^2 + 2 eBn + M^2} \pm \mu\,, \\ 
    &E^{\pm}_{p,\frac{1}{2}} = \sqrt{(\sqrt{p_z^2 + eBn + M^2} \pm \mu)^2 + \Delta^2 },
\end{align}
and correspond to the possible values of rotated charges, $|a|=0,1,\frac{1}{2}$ in the generic relation
\begin{equation}
E_{p,a}=\sqrt{{\bf p}_{\perp,a}^{2}+p_{z}^{2}+M^{2}},    
\end{equation}
where ${\bf p}_{\perp,a}^{2}=p_{x}^{2}+p_{y}^{2}$ for $a = 0$, and 
${\bf p}_{\perp,a}^{2}=2\left|a\right|eBn$ for $|a|=1,\frac{1}{2}$. The upper index in the dispersions of Eq.~\eqref{OmegaT} corresponds to the sign of the chemical potential in $E_{p,a}^{\pm} \equiv E_{p,a} \pm\mu$.
At zero temperature, which is the regime of interest in this work, the thermodynamic potential becomes
\begin{eqnarray}
\Omega_{T = 0} &=& \frac{\left(M-m_{c}\right)^{2}}{4G_{S}}+\frac{\Delta^{2}}{4G_{D}} - \Omega_0 - \Omega_1 - \Omega_{\frac{1}{2}}  
- \Omega_V\nonumber\\
\label{OmegaT0}
\end{eqnarray}
with the definitions
\begin{eqnarray}
\Omega_0 & = & 2\int\frac{d^{3}\vec{p}}{\left(2\pi\right)^{3}}\bigg[E_{p,0}+(\mu - E_{p,0})\theta(\mu - E_{p,0})\bigg],\nonumber\\
\Omega_1 & = & \frac{eB}{8\pi^{2}}\sum_{n=0}^{\infty}\alpha_{n}\int\limits_{-\infty}^{+\infty}dp_{z}\bigg[ 2E_{p,1} + (\mu - E_{p,1})\theta(\mu - E_{p,1})\bigg],\nonumber\\
\Omega_{\frac{1}{2}} & = & \frac{eB}{4\pi^{2}}\sum_{n=0}^{\infty}\alpha_{n}\int\limits_{-\infty}^{+\infty}dp_{z} \left(E_{p,\frac{1}{2}}^{+} + E_{p,\frac{1}{2}}^{-} \right),
\label{Omega_a}
\end{eqnarray} 
The last subtracted term, $\Omega_V$, in Eq.~\eqref{OmegaT0} is the value of $\Omega_{T = 0}$ at $\Delta=\mu = B=0 $ and $M = M_0$, which appears to make the pressure zero in the vacuum.

In the definitions above, terms multiplying the Heaviside theta functions come from the $T = 0$ limit of Eq.~\eqref{OmegaT}, while integrations over the dispersions are divergent and must be regularized.
In 2010, it became common in the literature to use simple step functions, $\theta(x-\Lambda)$, or to introduce smooth cutoffs, often implemented through form factors, to regularize the $B$-dependent integrals as~\cite{PhysRevD.82.045010,PhysRevD.83.025026,PhysRevLett.100.032007,Mandal:2012fq,Mandal:2016dzg,Mandal:2017ihr}
\begin{equation}
\sum_{n=0}^{\infty}\alpha_n\int\limits_{-\infty}^{+\infty}\frac{dp_z}{2\pi}\to
\sum_{n=0}^{\infty}\alpha_n\int\limits_{-\infty}^{+\infty}\frac{dp_z}{2\pi}U_{\Lambda}\left(\sqrt{p_z^2 + 2|a|eBn}\right),
\end{equation}
where $U\left(\sqrt{p_z^2 + 2|a|eBn}\right)$ is a smooth function that exhibits step-function-like behavior.
Nevertheless, it has been shown that in both cases, this procedure introduces strong unphysical oscillations in the behavior of different quantities as functions of the magnetic field~\cite{PhysRevD.99.116002}. Although the use of smooth regulators improves the results compared to the simple step function, the unphysical oscillations are often incorrectly associated with van Alphen--Haas (vA-dH) oscillations, even for parametrization sets where such oscillations should not be present~\cite{Allen:2015paa,Duarte:2015ppa,Duarte:2017nzv}. We will return to this point when discussing our numerical results. 

In recent years, the importance of separating medium effects from the divergent integrals prior to regularization has been widely discussed. Regarding the magnetic field contributions, in this work, we employ the magnetic field independent regularization (MFIR), following the approach previously adopted by~\cite{Allen:2015paa,Duarte:2015ppa,Avancini:2020xqe}. In the context of this work, the contributions  $|a| = 1$ and $|a| = 1/2$ 
to the thermodynamic potential become
(see~\cite{PhysRevC.79.035807,Allen:2015paa} for explicit details of MFIR implementation),
\begin{eqnarray}
\Omega_0 & = & 4\int_{\Lambda}\frac{d^{3}\vec{p}}{(2\pi)^{3}}\bigg[\sqrt{\vec{p}^{2}+M^{2}} - \sqrt{\vec{p}^{2}+M_0^{2}}\bigg] 
\nonumber\\
&+& 2\int\frac{d^{3}\vec{p}}{\left(2\pi\right)^{3}}\bigg[(\mu - E_{p,0})\theta(\mu - E_{p,0})\bigg],
\label{W0}
\end{eqnarray}
where the factor of 4 in front of the first integral in Eq.~\eqref{W0} differs from the factor of 2 in the first line of Eq.~\eqref{Omega_a} due to an algebraic manipulation performed when applying the MFIR. Specifically, the extra factor stems from the first term in $\Omega_1$ in the second line of Eq.~\eqref{Omega_a}. The remaining contribution of $\Omega_1$ reads,
\begin{eqnarray}
\Omega_1 &=& \frac{(eB)^2}{2\pi^2} \left[\zeta'( -1, \chi) + \frac{\chi - \chi^2}{2} \ln(\chi) + \frac{\chi^2}{4} \right] \nonumber \\
&+& \frac{eB}{4\pi^2} \sum_{n=0}^{p_{B,\text{max}}} \alpha_n \Biggl[ \mu \sqrt{\mu^2 - p_B^2}  \nonumber \\
& &  - p_B^2 \ln\bigg( \frac{\mu + \sqrt{\mu^2 - p_B^2}}{p_B} \bigg) \Biggr], 
\label{W1}
\end{eqnarray}

with $\chi = \frac{M^2}{2eB}$, $p_{B}=\sqrt{M^{2}+2eBn}$, and $p_{B,\text{max}}=\frac{\mu^{2}-M^{2}}{2eB}$, and 
\begin{eqnarray}
\Omega_{\frac{1}{2}} &=& 4\int_{\Lambda} \frac{d^{3}\vec{p}}{(2\pi)^{3}} \left[ \sqrt{\left( \sqrt{\vec{p}^{2} + M^{2}} + \mu \right)^{2} + \Delta^{2}}  \right. \nonumber\\
&& \left. + \sqrt{\left( \sqrt{\vec{p}^{2} + M^{2}} - \mu \right)^{2} + \Delta^{2}}   -2 \sqrt{\vec{p}^{2}+M_0^{2}}\right]\nonumber \\
& & + \frac{eB}{4\pi^{2}} \int_{-\infty}^{+\infty} dp_{z} F\left(p_{z}^{2}\right) \nonumber \\
& & + \frac{(eB)^{2}}{2\pi^{2}} \left[ \zeta^{\prime}\left(-1, x\right) + \frac{(x - x^{2})}{2} \ln(x) + \frac{x^{2}}{4} \right] \nonumber \\
& & + \frac{eB}{2\pi^{2}} \int_{-\infty}^{+\infty} dp_{z} \left[\sum_{n=1}^{\infty} F\left(p_{z}^{2} + neB\right) 
\right. \nonumber\\
&&\left. - \int_{0}^{\infty} dy~ F\left(p_{z}^{2} + eBy\right) \right], \label{W05}
\end{eqnarray}

where we have defined $x = \frac{M^2 + \Delta^2}{eB}$, and the function
\begin{eqnarray}
F(z)&=&\sqrt{\left(\sqrt{z+M^{2}}+\mu\right)^{2}+\Delta^{2}} 
\nonumber\\
&+&\sqrt{\left(\sqrt{z+M^{2}} -\mu\right)^{2}+\Delta^{2}} 
\nonumber\\
&-&2\sqrt{z+M^{2}+\Delta^{2}}.\label{Fz}
\end{eqnarray}

In Eqs.~\eqref{W1} and~\eqref{W05}, the symbol $\int_{\Lambda}$ indicates that the integrals are divergent and require some regularization scheme. It should be observed, however, that these integrals do not have an explicit dependence on the magnetic field, which is the central idea behind the MFIR scheme.

When using MFIR, we have divergent integrals of the type\footnote{Please refer to~\cite{Duarte:2018kfd} for the explicit form of the $\Delta$ and $M$ gap equations at zero magnetic field. The integral $I_{M,0}$ is the derivative of the first term in Eq.~\eqref{W0} with respect to $M$, while $I_{\Delta}$ and $I_{M,\frac{1}{2}}$ are the derivatives of the first line of Eq.~\eqref{W05} with respect to $\Delta$ and $M$, respectively. The extension to $eB\neq 0$ is straightforward.},
\begin{eqnarray}
I_{\Delta} = \int_{\Lambda}\frac{d^{3}\vec{p}}{\left(2\pi\right)^{3}}\left(\frac{1}{E_{\Delta}^+}+\frac{1}{E_{\Delta}^-}\right),
\label{Idelta}
\end{eqnarray}
with $E_{\Delta}^{\pm} = \sqrt{\left(\sqrt{\vec{p}^{2}+M^{2}}\pm\mu\right)^{2}+\Delta^{2}}$, and 
\begin{eqnarray}
I_{M,0} & = & \int_{\Lambda}\frac{d^{3}\vec{p}}{(2\pi)^{3}}\frac{1}{E_p},\\
I_{M,\frac{1}{2}} & = & \int_{\Lambda}\frac{d^{3}\vec{p}}{\left(2\pi\right)^{3}}\frac{1}{E_p}\left(\frac{E_p + \mu}{E_{\Delta}^+}+\frac{E_p - \mu}{E_{\Delta}^-}\right),
\end{eqnarray}
which do not depend explicitly on the magnetic field but exhibit both explicit and implicit dependence on the chemical potential $\mu$. Here, we defined $E_p = \sqrt{\vec{p}^{2}+M^{2}}$. The usual procedure, which we call the {\it traditional regularization scheme} (TRS), consists of introducing a sharp cutoff parameter $\Lambda$, which then becomes a model parameter together with the current quark mass $m_c$ and the scalar constant $G$.  
The MSS, on the other hand, completely removes the medium dependence from the divergent integrals by iteratively applying the identity
\begin{align}
&\frac{1}{a^2 + (E_p\pm\mu)^2 + \Delta^2} = \frac{1}{a^2 + \vec{p}^{\,2} + M_0^2} 
\nonumber\\
&\;\;\;\;\;\;\;\;+ \frac{M_0^2-(\Delta^2 + \mu^2 + M^2 \pm 2\mu E_p)}{(a^2 + \vec{p}^{\,2} + M_0^2)[a^2 + (E_p\pm\mu)^2 + \Delta^2]},
\label{ident}
\end{align}
for the $\Delta$ gap equation, for example. The integrand in Eq.~\eqref{Idelta} is obtained after integrating the left-hand side of Eq.~\eqref{ident} over $a$ from $-\infty$ to $+\infty$. The details of the MSS implementation are thoroughly discussed in Refs.~\cite{Farias:2005cr,Farias:2016let,Duarte:2018kfd,Avancini:2019ego}. Therefore, we restrict ourselves to presenting the final expressions here. For $\Delta$, we have
\begin{eqnarray}
I_{\Delta}^{\text{TRS}} & = & \int_0^{\Lambda}\frac{dp\,\vec{p}^2}{2\pi^{2}}\left(\frac{1}{E_{\Delta}^+}+\frac{1}{E_{\Delta}^-}\right),\label{Idtrs}\\
I_{\Delta}^{\text{MSS}} & = & 2I_{\text{quad}}(M_0) - (M^2-M_0^2+\Delta^2-2\mu^2)I_{\text{log}}(M_0)
\nonumber\\
&+& \left[\frac{3(A^2 + 4\mu^2M^2)}{4}-3\mu^2M_0^2\right]I_1 + 2I_2,\label{Idmss}
\end{eqnarray}
with $A = M_0^2 - \Delta^2 - \mu^2 - M^2$ and the definitions
\begin{align}
I_{\text{quad}}(M_0) &=  \int\frac{d^3\vec{p}}{(2\pi)^3}\frac{1}{\sqrt{\vec{p}^2 + M_0^2}}\label{Iquad}\;,\\
I_{\text{log}}(M_0) &= \int\frac{d^3\vec{p}}{(2\pi)^3}\frac{1}{(\vec{p}^2 + M_0^2)^{\frac{3}{2}}}\label{Ilog}\;,\\
I_1  &=  \int\frac{d^3\vec{p}}{(2\pi)^3}\frac{1}{(\vec{p}^2 + M_0^2)^{\frac{5}{2}}}\;,\\
I_2 &= \frac{15}{32}\sum_{j= \pm 1}\int\frac{d^3\vec{p}}{(2\pi)^3}\Biggl\{\int\limits_0^1 dt (1-t)^2
\nonumber\\
&\times \frac{(A - 2j\mu E_p)^3}{\left[(2j\mu E_p - A)t + \vec{p}^2 + M_0^2\right]^{\frac{7}{2}}}\Biggr\}.
\end{align}
Similarly, for the $M$ gap equation, we have
\begin{eqnarray}
I_{M,0}^{\text{TRS}} & = & \int\limits_0^{\Lambda}\frac{dp\,\vec{p}^2}{2\pi^{2}}\frac{1}{E_p},\\
I_{M,\frac{1}{2}}^{\text{TRS}} & = & \int\limits_0^{\Lambda}\frac{dp\,\vec{p}^2}{2\pi^{2}}\frac{1}{E_p}\left(\frac{E_p + \mu}{E_{\Delta}^+}+\frac{E_p - \mu}{E_{\Delta}^-}\right)
\end{eqnarray}
and 
\begin{align}
&I_{M,0}^{\text{MSS}}  =  I_{\text{quad}}(M_0)+\frac{M_{0}^{2}-M^{2}}{2}I_{\text{log}}(M_0)+I_3,\\
&I_{M,\frac{1}{2}}^{\text{MSS}} = 2I_{\text{quad}}(M_0) - (M^2 - M_0^2 + \Delta^2)I_{\text{log}}(M_0) \nonumber\\
&\;\;+ 3\left[\frac{A^2}{4} + \mu^2(M^2 - M_0^2 - A)\right]I_1 + 2I_2 + I_4,
\end{align}
with the remaining definitions,
\begin{align}
I_3 & = \frac{3}{4}\int\frac{d^{3}\vec{p}}{(2\pi)^{3}} 
\int\limits_{0}^{\infty}
\frac{t\,dt}{\sqrt{1+t}}\frac{(M_{0}^{2}-M^{2})^{2}}{\left[(p^{2}+M_{0}^{2})t+p^{2}+M^{2}\right]^{\frac{5}{2}}}\;,\\
I_4  &=  \frac{15}{16}\sum_{j = \pm 1}\int\frac{d^3\vec{p}}{(2\pi)^3}\Biggl\{ \int\limits_0^{\infty}\frac{t^2 dt}{\sqrt{1+t}}\nonumber\\ 
&\times\frac{j\mu(A - 2j\mu E_p)^3}
{E_p\left[(\vec{p}^2 + M_0^2)t + (E_p + j\mu)^2 + \Delta^2\right]^{\frac{7}{2}}}\Biggr\}.
\end{align}

In the MSS framework, the thermodynamic potential becomes, after collecting the regularized integrals in both the first line of~\eqref{W0} and the first and second lines of~\eqref{W05},
\begin{align}
4 \int_{\Lambda}& \frac{d^3\vec{p}}{(2\pi)^3}\Biggl[E_p + E_{\Delta}^+ + E_{\Delta}^-  -3E_0 \Biggl]\nonumber\\ 
&= 2(M^2 - M^2_0 + 2\bar{M})I_{\text{quad}}(M_0)\nonumber\\ 
&- \bigg[ \frac{(M^2 -M^2_0 )^2}{2} + \bar{M}^2 - 4\Delta^2 \mu^2\bigg] I_{\text{log}}(M_0) \nonumber\\ 
&+ 4 \int \frac{d^3 \vec{p}}{(2\pi)^3} \bigg \{  \frac{(M^2 -M^2_0)^2 + 2\bar{M}^2 - 8 \Delta^2 \mu^2}{8E^3_0}  
\nonumber\\
&- \frac{M^2 -M^2_0 + 2 \bar{M}}{2E_0}  
+ E_p + E^{+}_{\Delta} + E^{-}_{\Delta} - 3E_0\bigg\}\nonumber\\
\end{align}
with the extras definitions $E_0 = \sqrt{\vec{p}^2 + M_0^2}$ and $\Bar{M} = \Delta^2 + M^2 - M^2_0$.\\

In this work, two separation schemes are employed; MFIR, which isolates purely magnetic contributions, and MSS, which isolates density contributions. Throughout the paper, MFIR refers solely to the magnetic separation, while the full disentanglement of medium effects is achieved through the combined implementation of MFIR and MSS.
In this section, we presented the expressions for different regularization methods, along with references for readers interested in the detailed derivations. In the subsequent section, we present a comparative analysis of these methods through numerical results, followed by a discussion of their implications for physical quantities.  

\section{Numerical Results}
\label{Sec:results}

In order to discuss the numerical results, it is necessary to establish the model parameters, specifically the current quark mass $m_c$, the scalar coupling $G_S$, and the previously mentioned energy scale $\Lambda$. These parameters are determined by fitting them to empirical values; the pion decay constant $f_{\pi} = 93.2$ MeV, the pion mass $m_{\pi} = 135$ MeV, and the quark condensate in the vacuum, $\langle \bar{q} q \rangle ^{1/3}_0 = -250$ MeV. This procedure yields $G_S = 4.75$ GeV$^{-2}$, $m_c = 4.99$ MeV, and $\Lambda = 660$ MeV. For all the regularization schemes, the vacuum quark mass (which is also the mass scale for MSS) is $M_0 \sim 302$ MeV, and the diquark coupling constant $G_D = 0.75G_S$ is obtained via a Fierz transformation~\cite{Buballa:2003qv}. Additionally, a smoothness parameter $\alpha = 0.01 \Lambda$ is employed for the smooth functions of the Fermi-Dirac form, as in~\cite{Mandal:2017ihr}:
\begin{eqnarray}
U(x) =\frac{1}{2}\left[1 - \tanh\left(\frac{x/\Lambda - 1}{\alpha}\right)\right].
\end{eqnarray}
We call this scheme nonmagnetic field independent regularization (nMFIR), following the notation of~\cite{PhysRevD.99.116002}.
 
\subsection{The finite magnetic field case} 

\begin{center}
 \begin{figure}[htpb!]
 {\includegraphics[scale=0.33]{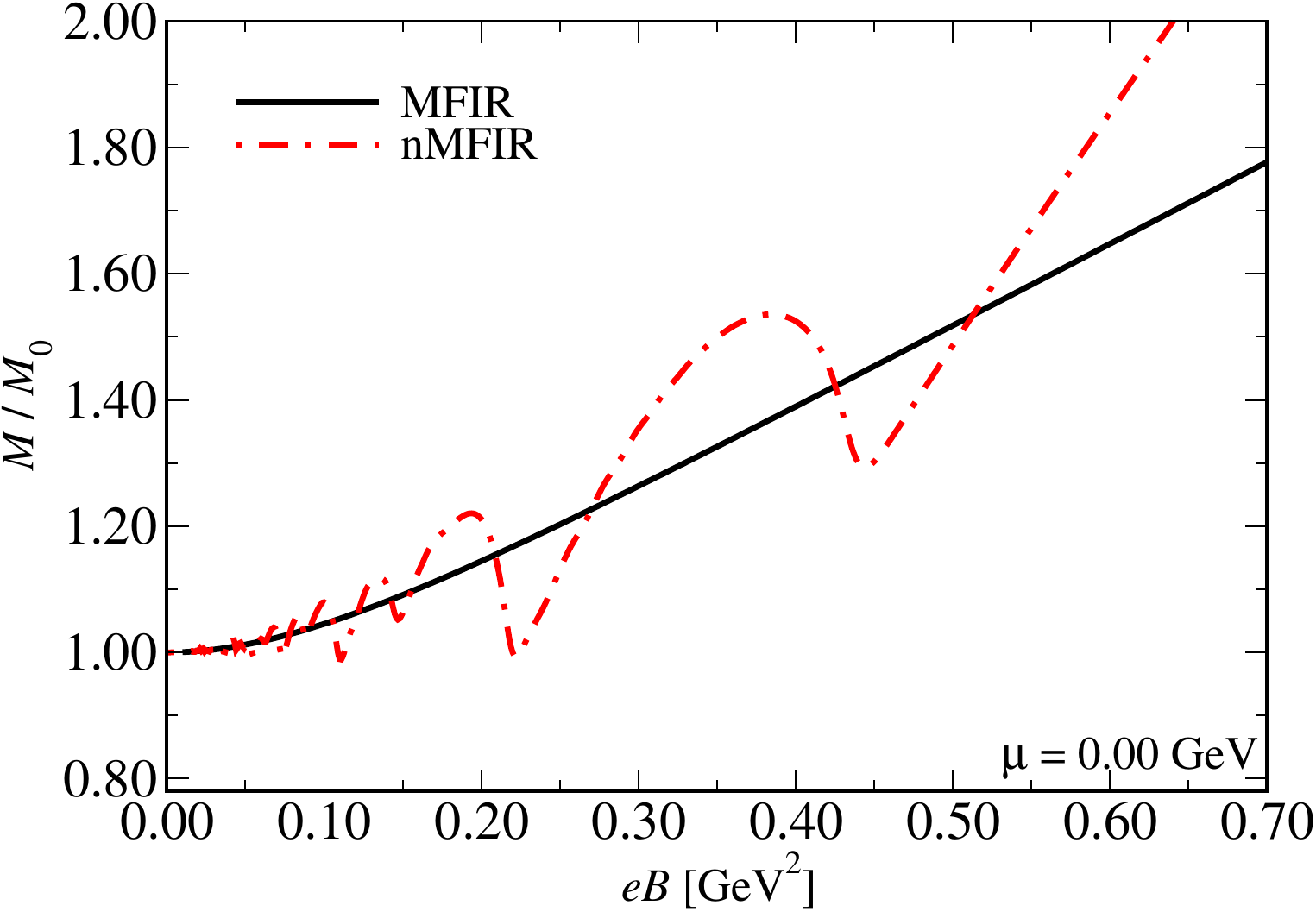}}
  \caption{Normalized quark mass
 $M / M_{0}$ as a function of $eB$ for $G_D/G_S = 0.0$, comparing MFIR and nMFIR.}
  \label{Fig1}
 \end{figure}
\end{center}

Figure~\ref{Fig1} compares the normalized effective quark mass $M/M_0$ as a function of the magnetic field at zero chemical potential (and, consequently, $\Delta = 0$). It is well-established that the chiral condensate and the constituent quark mass, given their linear relation, increase with the magnetic field, a phenomenon known as magnetic catalysis~\cite{Gusynin:1995nb}. As observed in this figure, both regularizations exhibit this phenomenon; however, the nMFIR scheme introduces strong unphysical oscillations that are more pronounced as the magnetic field increases, in contrast to the behavior obtained for MFIR~\cite{PhysRevC.79.035807}. These artifacts are often mistaken for the well-known (vA-dH) oscillations, which are associated with discontinuities in the density, and are present at finite $\mu$, as can be seen in the following results. 

The analysis of the thermodynamic potential shows that the oscillations originate solely from the $\Omega_{1}$ term, which contains a theta function dependent on the magnetic field, thereby establishing an upper limit for the Landau-level summation. However, it notably vanishes at zero chemical potential. While $\Omega_{0}$ is independent of $eB$, the coupling of quark species to $\Delta$ in $\Omega_{\frac{1}{2}}$ eliminates the theta function in the $T = 0$ limit. Consequently, for nonzero values of $\Delta$, the density is consistently nonzero across all Landau levels. This is the case for the $N_c = 2$, where there are no theta functions in the zero temperature limit, and all the oscillations are strictly unphysical~\cite{Duarte:2015ppa}. 

\begin{center}
   \begin{figure}[htpb!]
 \subfigure[]{\includegraphics[scale=0.33]{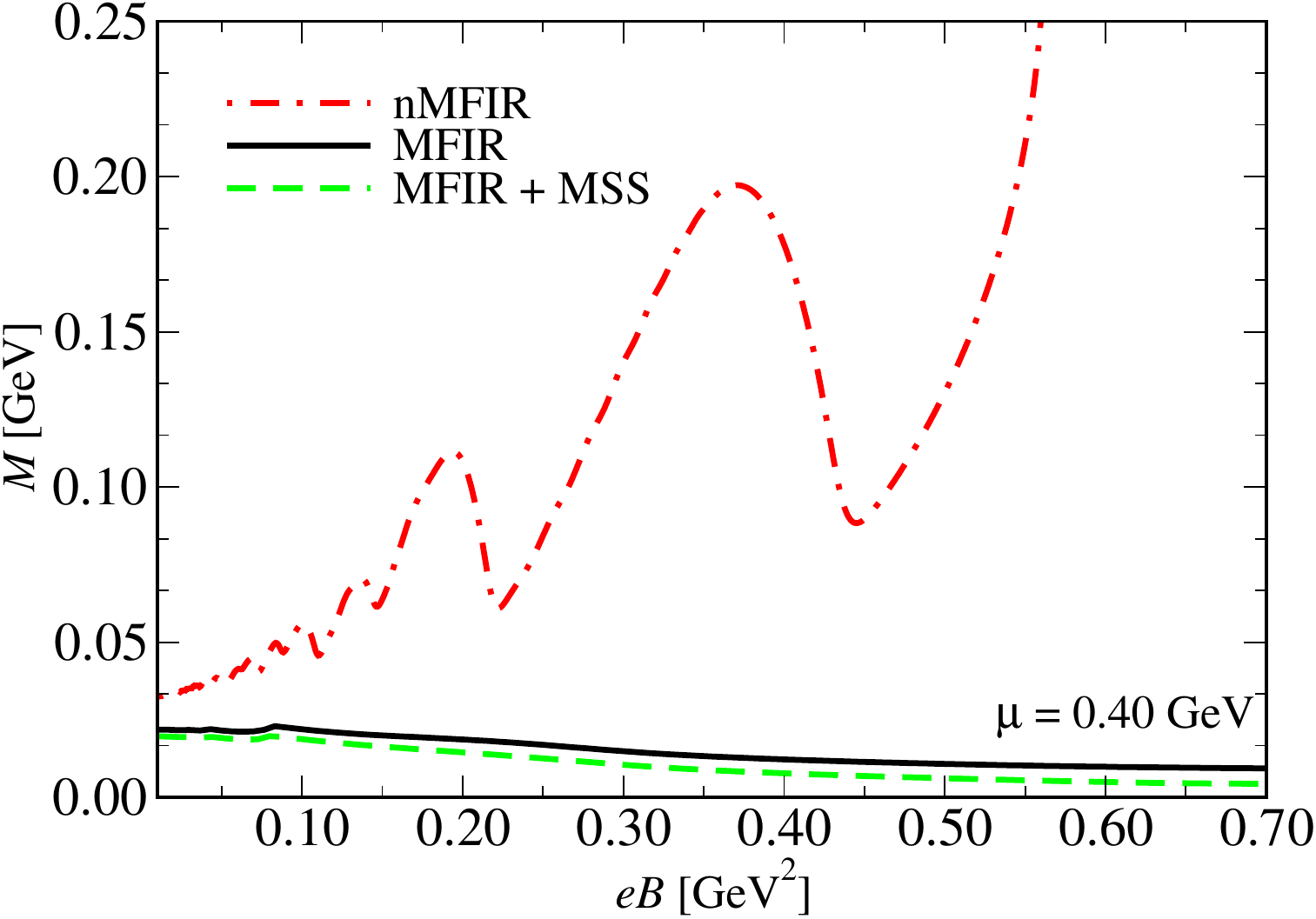}}
  \subfigure[]{\includegraphics[scale=0.33]{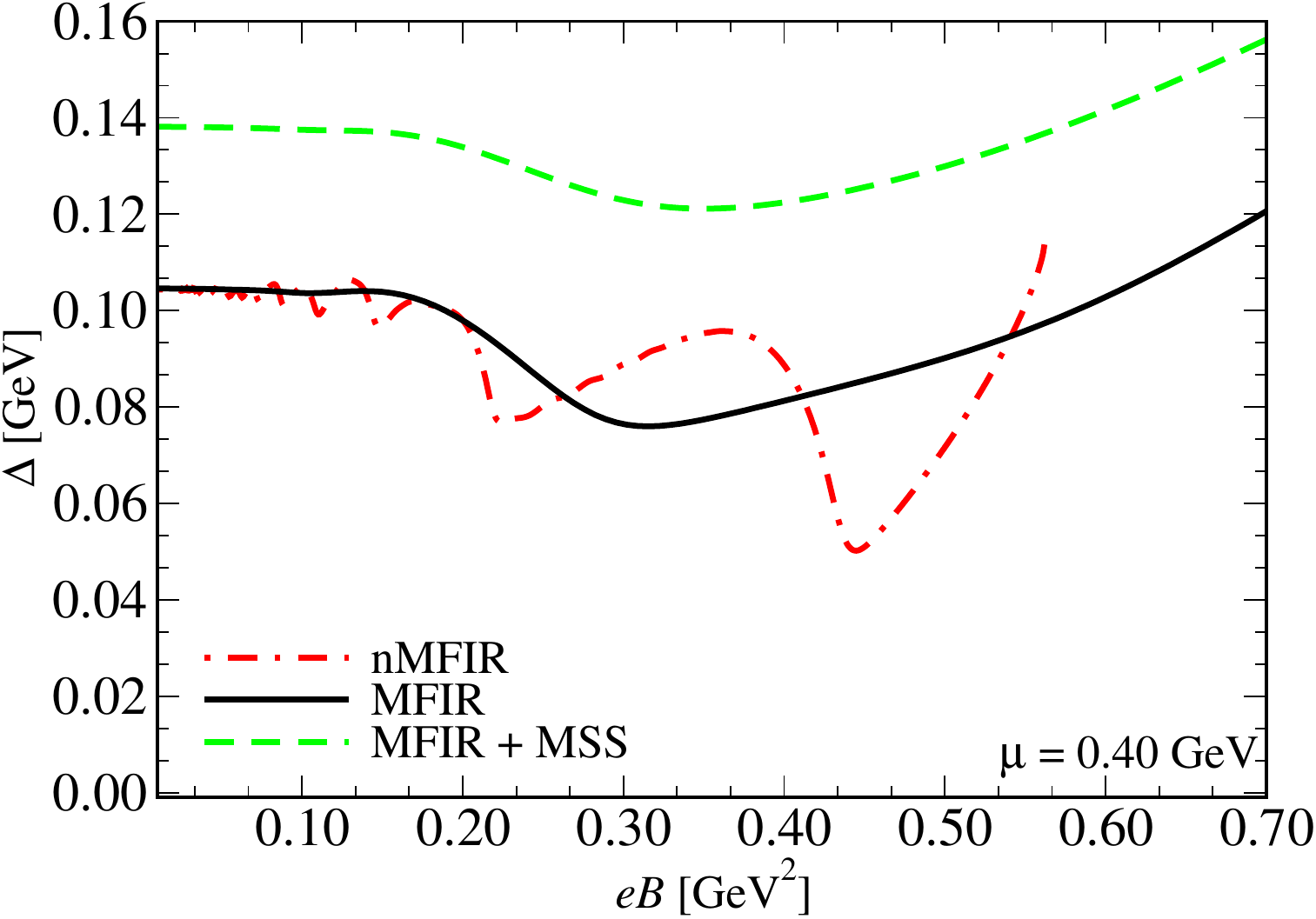}}\\ 
  \caption{Constituent quark mass 
$M$ [panel (a)] and diquark condensate $\Delta$ [panel (b)] as functions of the magnetic field, comparing nMFIR, MFIR, and MFIR + MSS.}
  \label{Fig2}
  \end{figure}
  \end{center}

Both panels in Fig.~\ref{Fig2} compare the order parameters obtained from the nMFIR, MFIR, and MSS methods as a function of the magnetic field, at $\mu$ = 0.4 GeV. Notably, while vAdH oscillations are evident in the MFIR results, the trends diverge significantly between the methods. In the MFIR and MFIR + MSS approaches, both the diquark condensate $\Delta$ and the constituent quark mass $M$ decrease with increasing magnetic field, demonstrating inverse magnetic catalysis (IMC) accompanied by smooth oscillations. Conversely, the nMFIR scheme exhibits an increase in both $\Delta$ and $M$ with magnetic field strength, accompanied by pronounced unphysical oscillations.

\subsection{The finite chemical potential case} 

\begin{center}
\begin{figure}[htpb!]
 \subfigure[]{\includegraphics[scale=0.33]{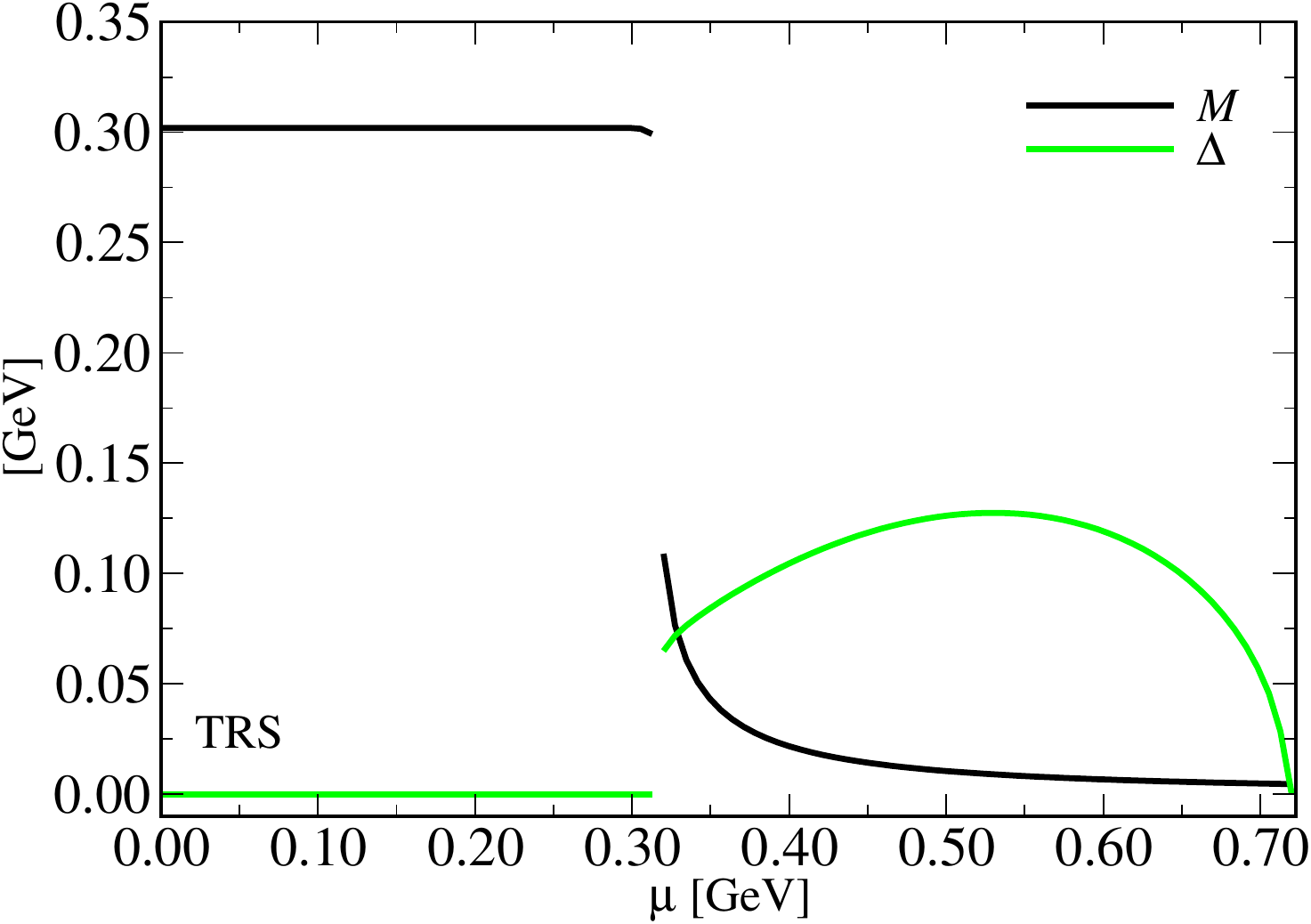}} 
  \subfigure[]{\includegraphics[scale=0.33]{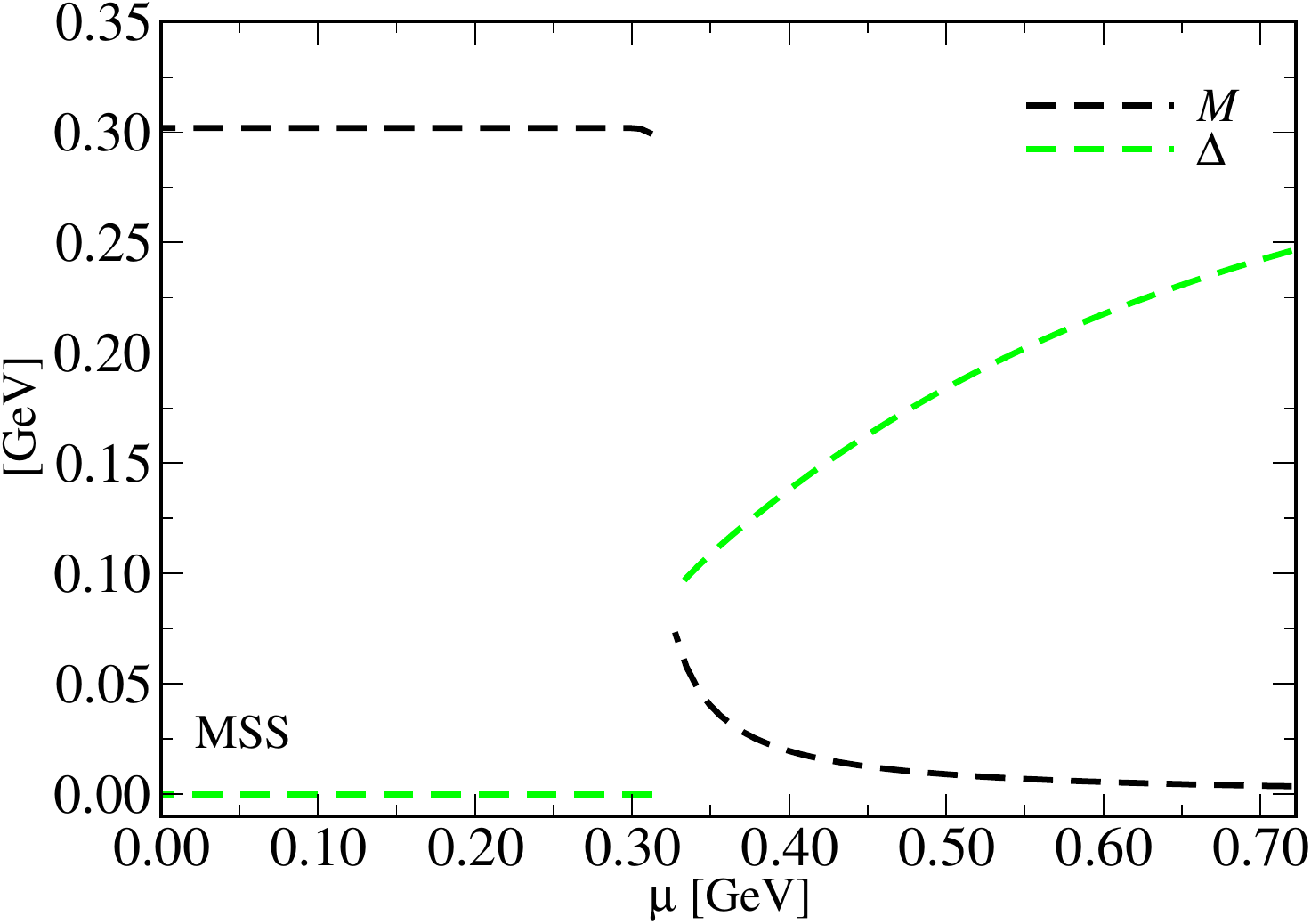}}
  \caption{Constituent quark mass $M$ and diquark condensate $\Delta$ as functions of the chemical potential, comparing TRS [panel (a)]  and MSS [panel (b)] at $eB$ = 0.}
   \label{Fig3}
  \end{figure}
 \end{center}

This section presents the results obtained by varying the quark chemical potential at fixed magnetic field values. Since we are interested in the behavior of the physical quantities when the density increases, we compare the TRS and MSS results using MFIR in both cases. The panels in Fig.~\ref{Fig3} illustrate the effective quark mass and the diquark condensate within the TRS [panel (a)] and MSS [panel (b)] frameworks at zero magnetic field. The effective quark mass $M$ exhibits qualitatively similar behavior across both methods. However, the diquark condensate reveals a significant discrepancy in the high chemical potential region. 

 In the TRS approach, the diquark condensate initially increases with $\mu$, and at a value smaller than the model scale $\Lambda$ it starts to decrease until it reaches zero. This contrasts with the MSS approach, where the condensate shows a steady increase at high chemical potential. This behavior in the MSS approach provides a more realistic description of the diquark condensate in the high chemical potential regime compared to other methods, such as lattice QCD simulations with $N_c = 2$ ~\cite{Kogut:2001na, Braguta:2016cpw}, chiral perturbation theory ~\cite{Kogut:2000ek}, and recent renormalization group treatment~\cite{Gholami:2024diy}, all of which demonstrate a similar trend for $\Delta$.

\begin{center}
\begin{figure}[htpb!]
 \includegraphics[scale=0.33]{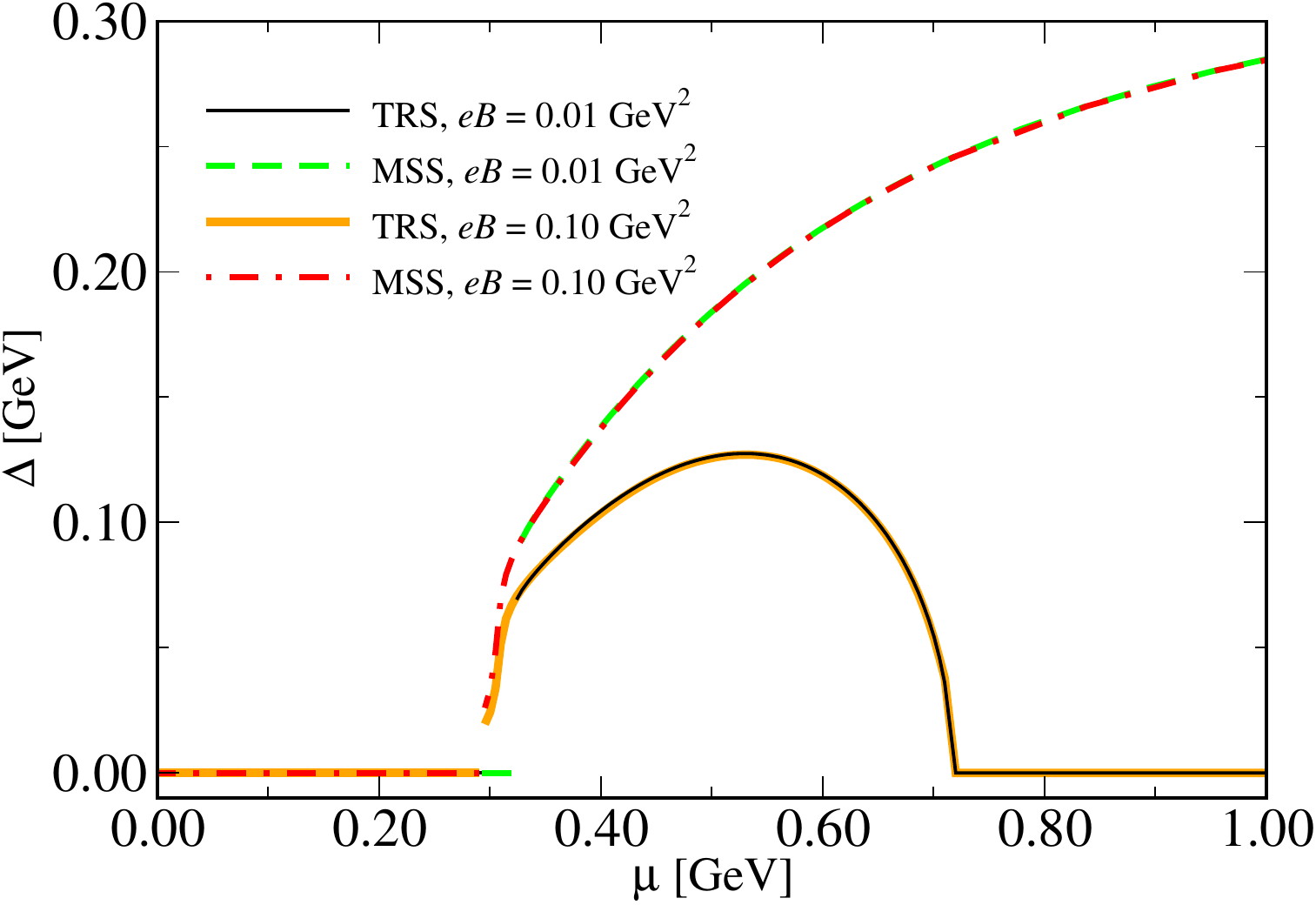}
  \caption{Diquark condensate $\Delta$, as a function of $\mu$, comparing MFIR and MFIR + MSS at finite magnetic field.}
   \label{Fig4}
  \end{figure}
 \end{center}

\begin{center}
\begin{figure*}[htpb!]
\subfigure[]{\includegraphics[scale=0.33]{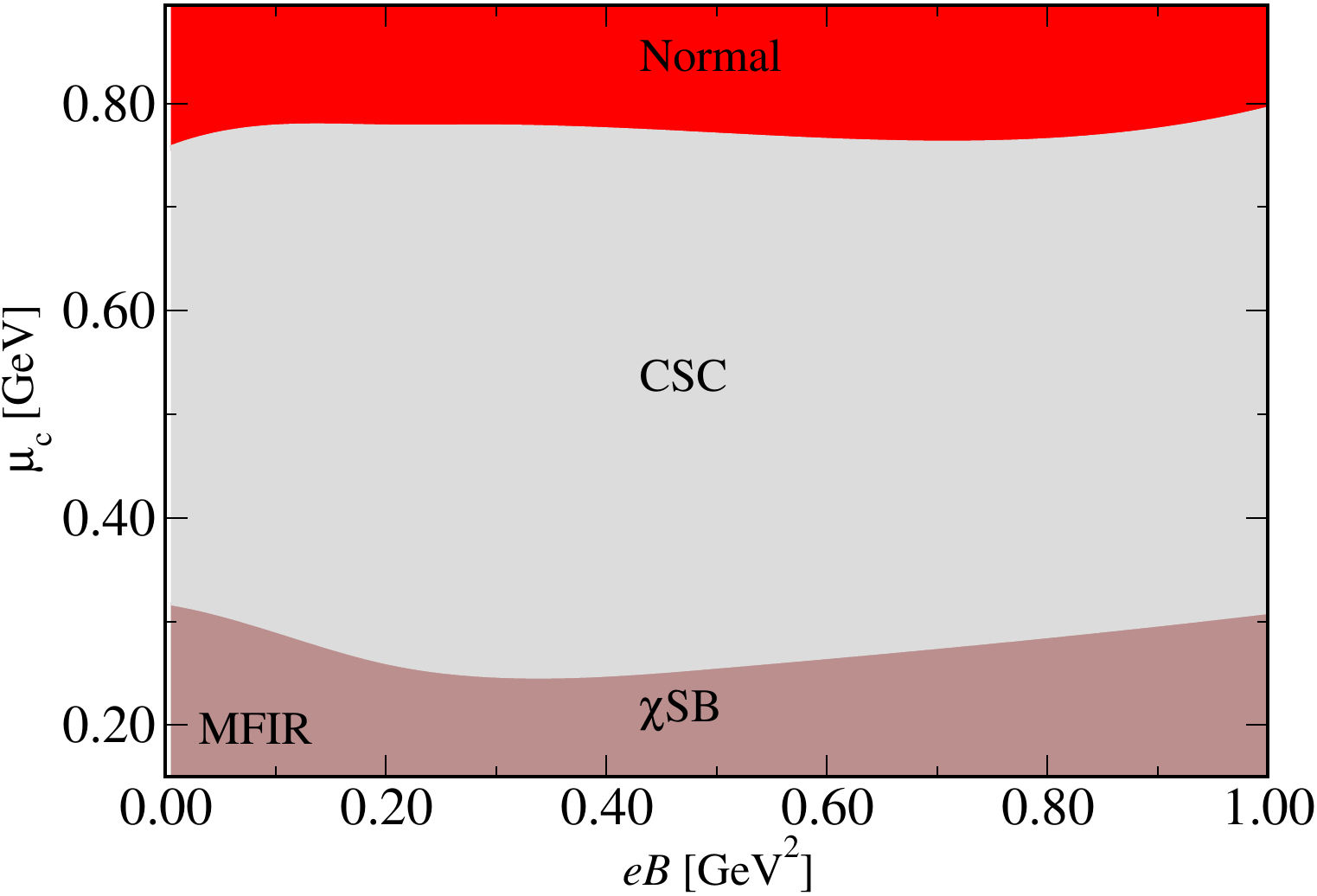}}
\subfigure[]{\includegraphics[scale=0.33]{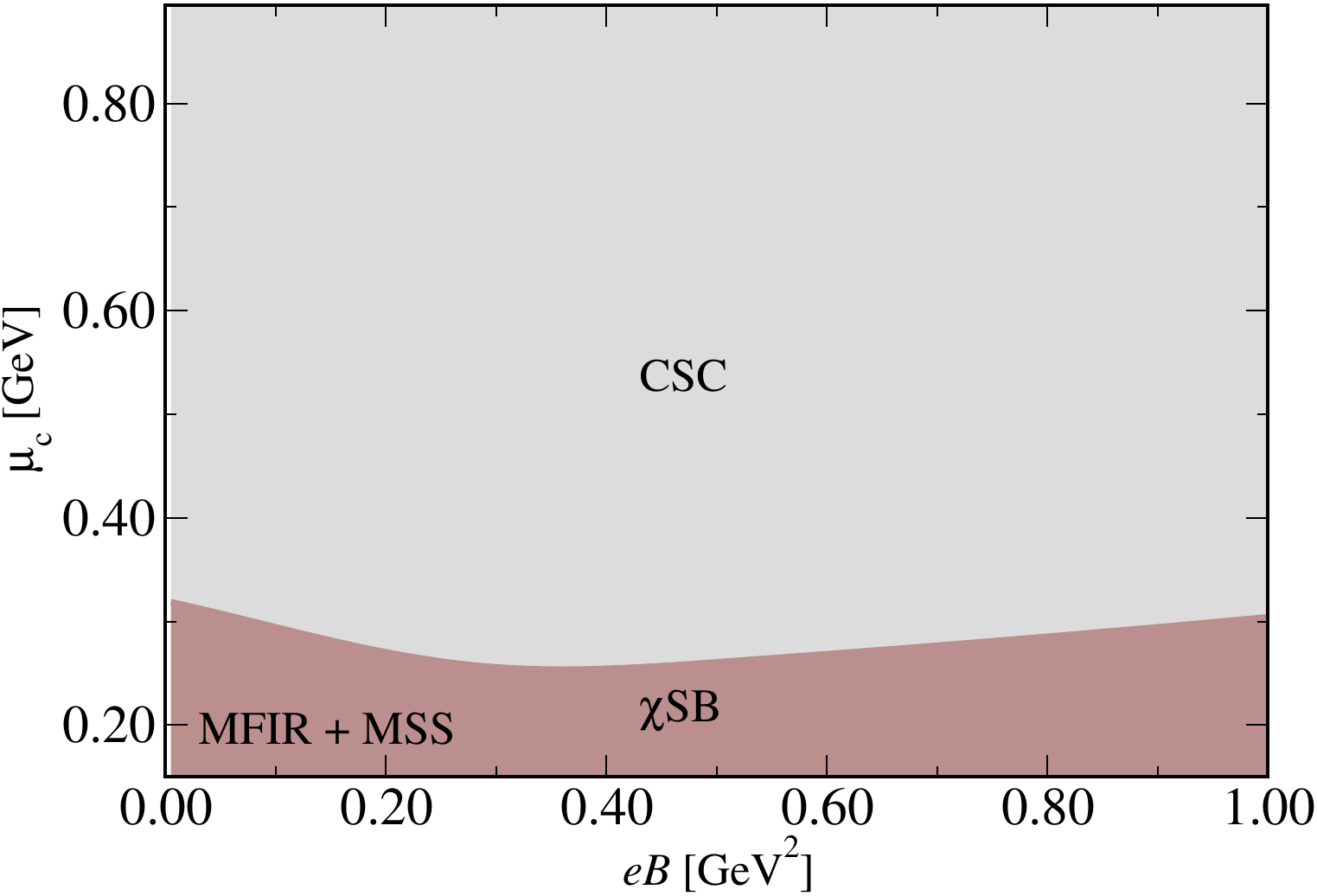}}
  \caption{Phase diagram in the $eB \times \mu$ plane, comparing the MFIR method in panel (a) and the MFIR + MSS method in panel (b). These diagrams display the chiral symmetry breaking phase ($\chi SB$), the color superconducting phase (CSC), and a phase with restored chiral symmetry (Normal).}
   \label{Fig5}
  \end{figure*}
 \end{center}

Figure~\ref{Fig4} illustrates the diquark condensate as a function of the chemical potential $\mu$ at finite magnetic field. We present only the $\Delta$ results, as the constituent quark mass exhibits the same qualitative behavior for both methods. This is consistent with Fig.~\ref{Fig3} at $eB = 0$, and this qualitative similarity persists for the finite values of $eB$ considered in this work. These results demonstrate that the behavior for $\Delta$ remains consistent with the $eB = 0$ case; $\Delta$  increases monotonically with $\mu$ for MSS, while for TRS, it initially increases but eventually decreases to zero. 
A notable difference compared to Fig.~\ref{Fig3} is the value of $\mu$ at which the diquark condensate is formed and vanishes in the TRS case. This difference is better visualized in Fig.~\ref{Fig5}, where we present the results for the phase diagram on the $\mu\times eB$ plane. 

This figure highlights the importance of the proper divergence treatment in the model when medium effects are included. The left panel shows that TRS predicts a normal phase with chiral symmetry restored and no color superconducting phase at large values of $\mu$. This prediction is an artifact of improper regularization, conflicting with other established approaches and with the expected physical picture at high chemical potential and low temperature. In this regime, the system is susceptible to the Cooper instability,\footnote{See~\cite{PhysRev.104.1189} for details on this instability and the Cooper theorem for Fermi systems. The analogous case for QCD was proposed as early as  1975~\cite{PhysRevLett.34.1353}.} leading to the formation of Cooper pairs as the most stable state for deconfined quarks.

\subsection{Thermodynamic quantities}

In this section, we present some thermodynamic quantities of the system, focusing on how the thermodynamics of the system is affected by the presence of an external magnetic field.  
An important outcome of this analysis is the emergence of anisotropy; in the presence of a strong magnetic field, the pressure splits into parallel and transverse components~\cite{Ferrer:2010wz,Ferrer:2020tlz}.
The component of the pressure parallel to the direction of the magnetic field, $p^{\parallel}$, is obtained directly from the thermodynamic potential given in Eq.~\eqref{OmegaT0}, in the same way as done at $B = 0$,
\begin{equation}
p^{\parallel}(M,\mu,eB) = -\Omega_{T=0}(M,\mu, eB).
\label{Pressure}
\end{equation}
To guarantee that this component vanishes at $\mu = 0$, we subtract the vacuum contribution, $p_V^{\parallel} = p^{\parallel}(M_0,0,eB)$, and introduce the normalized parallel pressure, $p_N^{\parallel}$, as
\begin{equation}
p_N^{\parallel} = p^{\parallel} - p_V^{\parallel},
\label{PparN}
\end{equation}
where, for simplicity, we omit the functional dependencies of all thermodynamic quantities from this point onward. It is important to note that the definition~\eqref{Pressure} would include a purely magnetic contribution, $-\frac{B^2}{2}$, but in anticipation of the normalization introduced in~\eqref{PparN} we have omitted this term. 

The magnetization,  
{\termchancery M },
is given by the derivative of the thermodynamic potential with respect to the magnetic field and can be expressed in terms of the parallel pressure as
\begin{equation}
\text{{\termchancery M} } = \left.\frac{\partial p_N^{\parallel}}{\partial B}\right|_{\mu},
\label{magnetization}
\end{equation}
with the derivative evaluated at fixed chemical potential 
$\mu$. From these definitions, the energy density follows as
\begin{equation}
\varepsilon_N = -p_N^{\parallel} + \mu_B n_B + \text{{\termchancery M} }B,
\label{epsilon}
\end{equation}
where the baryon number density, $n_B$, can be obtained as the derivative of the pressure with respect to the baryon chemical potential, $\mu_B = N_c\mu$, at fixed magnetic field,
\begin{equation}
n_B = \left.\frac{\partial p_N^{\parallel}}{\partial \mu_B}\right|_{eB}.
\label{rho}
\end{equation}

Finally, the speed of sound can also be decomposed into two distinct components at fixed magnetic field~\cite{Ferrer:2022afu}: one along the direction parallel to the field, 

\begin{equation}
(c_s^\parallel)^2 = \left.\frac{d p_N^{\parallel}}{d \varepsilon_N}\right|_{eB},
\label{cs2parallel}
\end{equation}
and another in the direction perpendicular to the field, 

\begin{equation}
(c_s^\perp)^2= \left.\frac{d p_N^{\perp}}{d \varepsilon_N}\right|_{eB},
\label{cs2perp}
\end{equation}
where the perpendicular component of the pressure is given in terms of the magnetization and the parallel component as $p^{\perp}_N = p_N^{\parallel} - \text{\termchancery M} \ B$.

\begin{figure}[htpb!]
\subfigure[]{\includegraphics[scale=0.33]{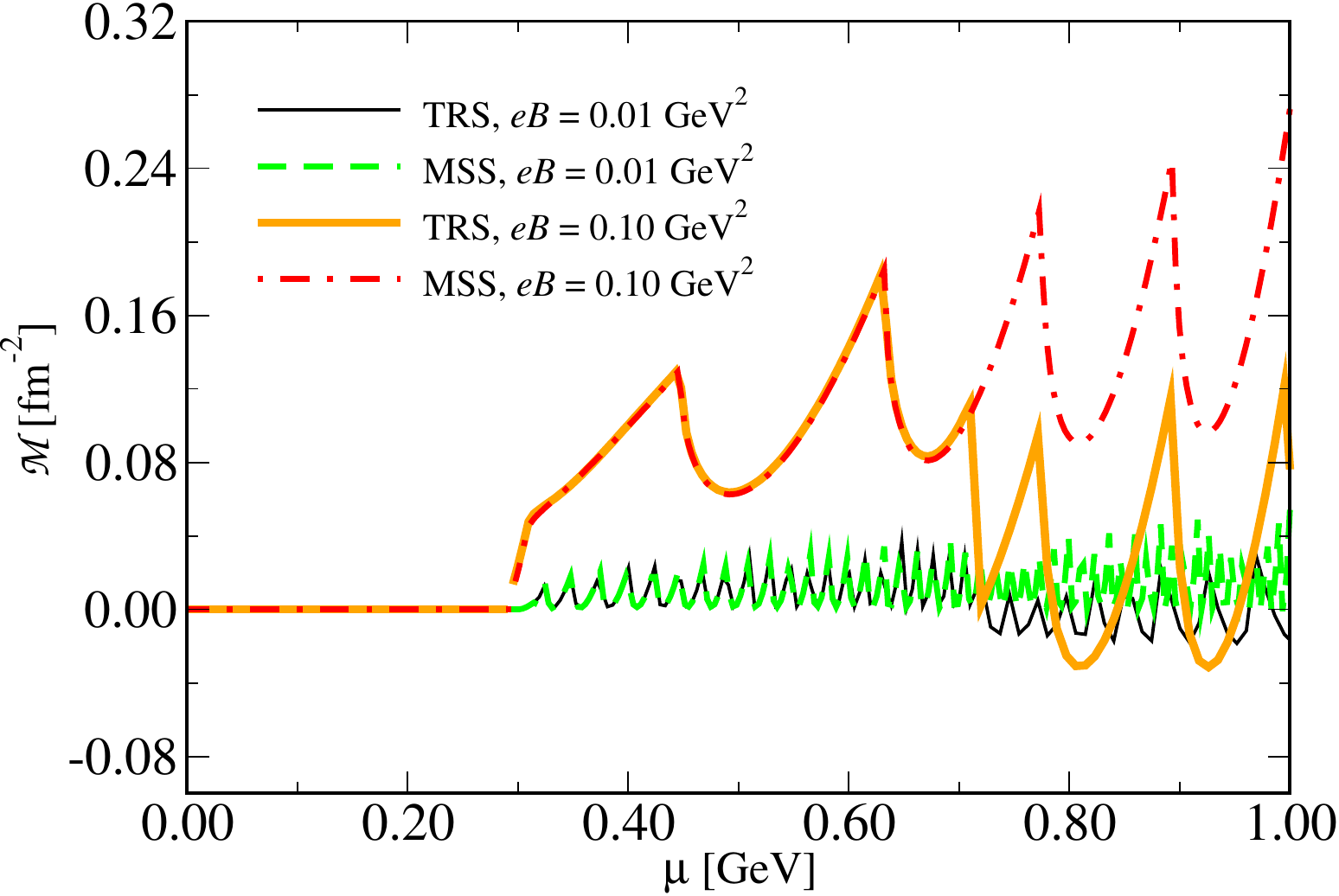}}
\subfigure[]{\includegraphics[scale=0.33]{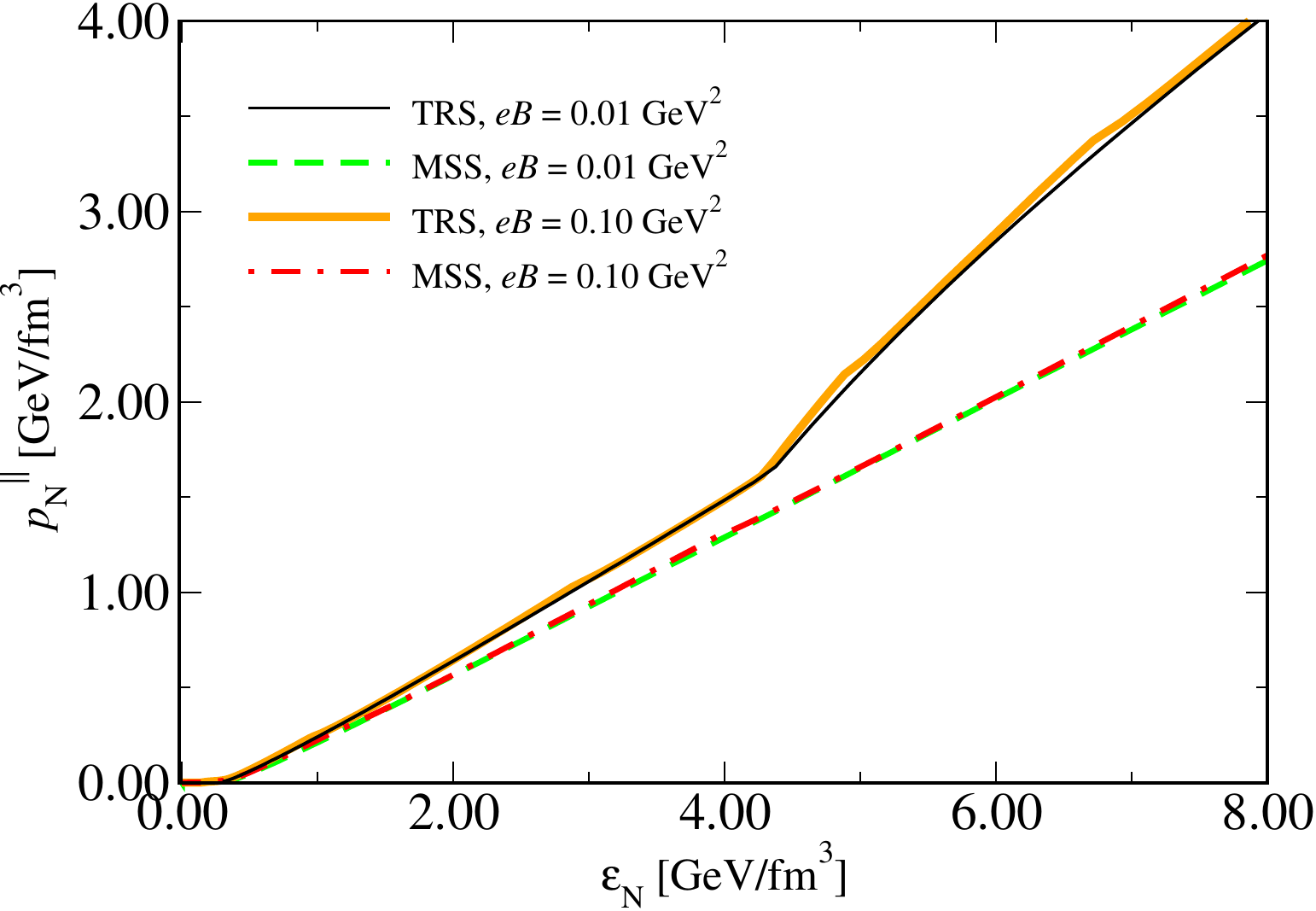}} 
  \caption{Magnetization {\termchancery M } as a function of chemical potential $\mu$ [panel (a)] and equation of state $p_N^{\parallel} \times \varepsilon_N$ [panel (b)] for a finite magnetic field.}
   \label{Fig6}
\end{figure}

The figures in this section present the results for thermodynamic quantities for $eB = 0.01$ and 0.1 GeV$^2$, which correspond to $eB\approx 5\times 10^{17}$ G and $5\times 10^{18}$ G, respectively. The results obtained with TRS and MSS are compared, both implemented within the MFIR scheme. Panel (a) of Fig.~\ref{Fig6} displays the magnetization, $\text{\termchancery M}$, as a function of the chemical potential $\mu$. For the weaker field, both the TRS and MSS exhibit pronounced oscillations with increasing $\mu$, a behavior that can be associated with the filling of Landau levels. 
At $eB = 0.10~\text{GeV}^2$ and nonvanishing diquark pairing, both regularization schemes converge toward similar behaviors, as Landau quantization dominates.
At higher values of chemical potential, however, when $\Delta$ begins to decrease, their predictions differ: while TRS yields an eventually negative magnetization, the MSS scheme maintains a consistently positive magnetization across the entire range of $eB$ values considered, signaling a robust paramagnetic response of dense matter. 
The qualitative dependence of the magnetization on the chemical potential is also compatible with results obtained for cold magnetized quark matter within the nonlocal Nambu--Jona-Lasinio model~\cite{Ferraris:2025fva}, suggesting that the main physical features remain stable across different effective descriptions.
Although the diquark condensate vanishes within TRS at chemical potentials beyond the model scale, recent studies show that the MSS can be consistently extrapolated to $\mu > \Lambda$ in different contexts, yielding results in agreement with alternative approaches, including lattice simulations~\cite{Lopes:2025rvn} and perturbative QCD calculations~\cite{Pasqualotto:2025kpo,sym18020220}.
In panel (b) of Fig.~\ref{Fig6}, one can see the equation of state $p_N^{\parallel}\times \varepsilon_N$. The equation of state obtained within MSS is systematically softer than that produced by TRS, reflecting the proper separation of medium contributions without introducing artificial cutoffs. By preserving the physics near the Fermi surface, the MSS approach provides a more reliable description at high densities, with direct relevance for astrophysical applications, particularly the modeling of compact stars.

\begin{figure}[htpb!]
\subfigure[]{\includegraphics[scale=0.33]{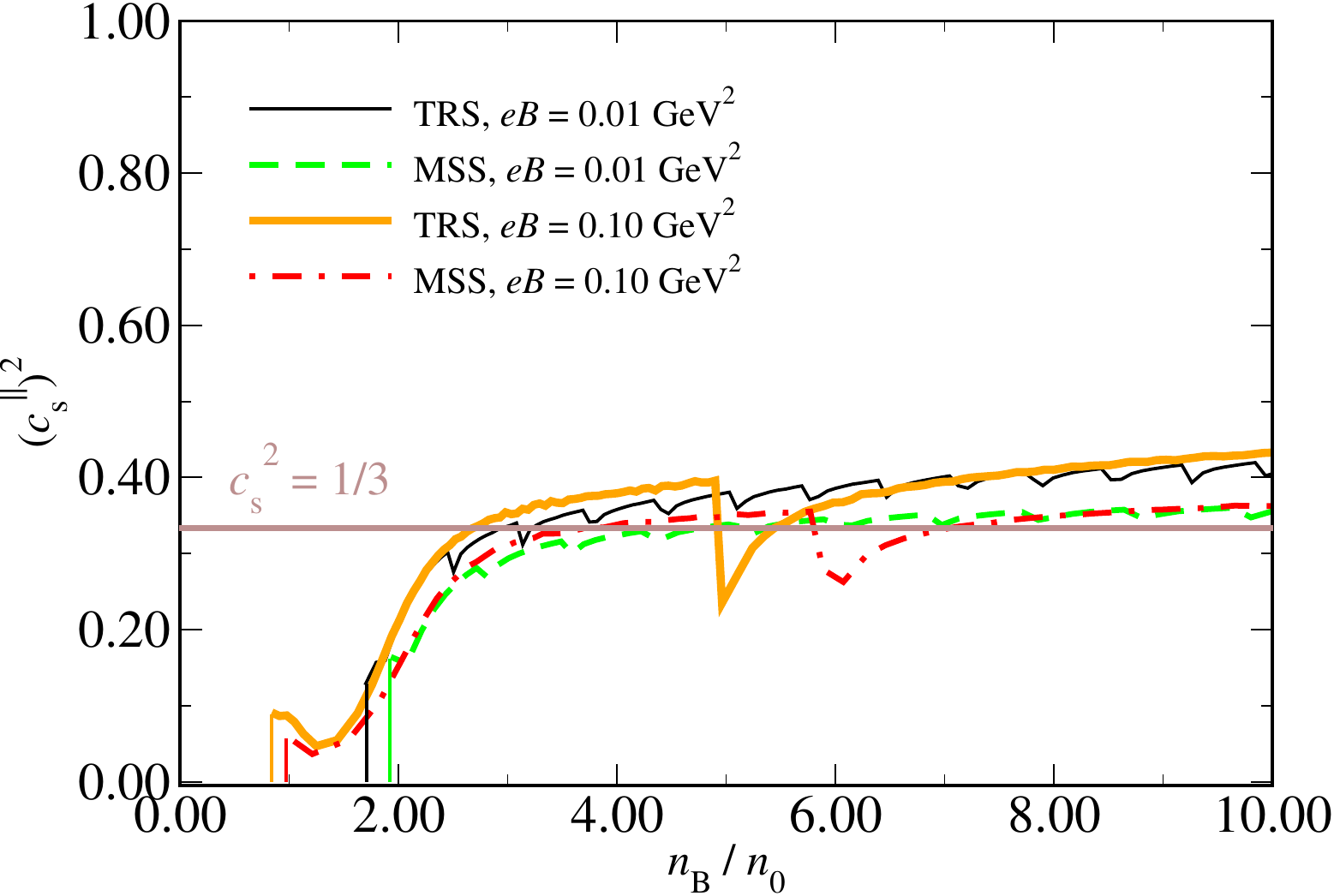}}
\subfigure[]{\includegraphics[scale=0.33]{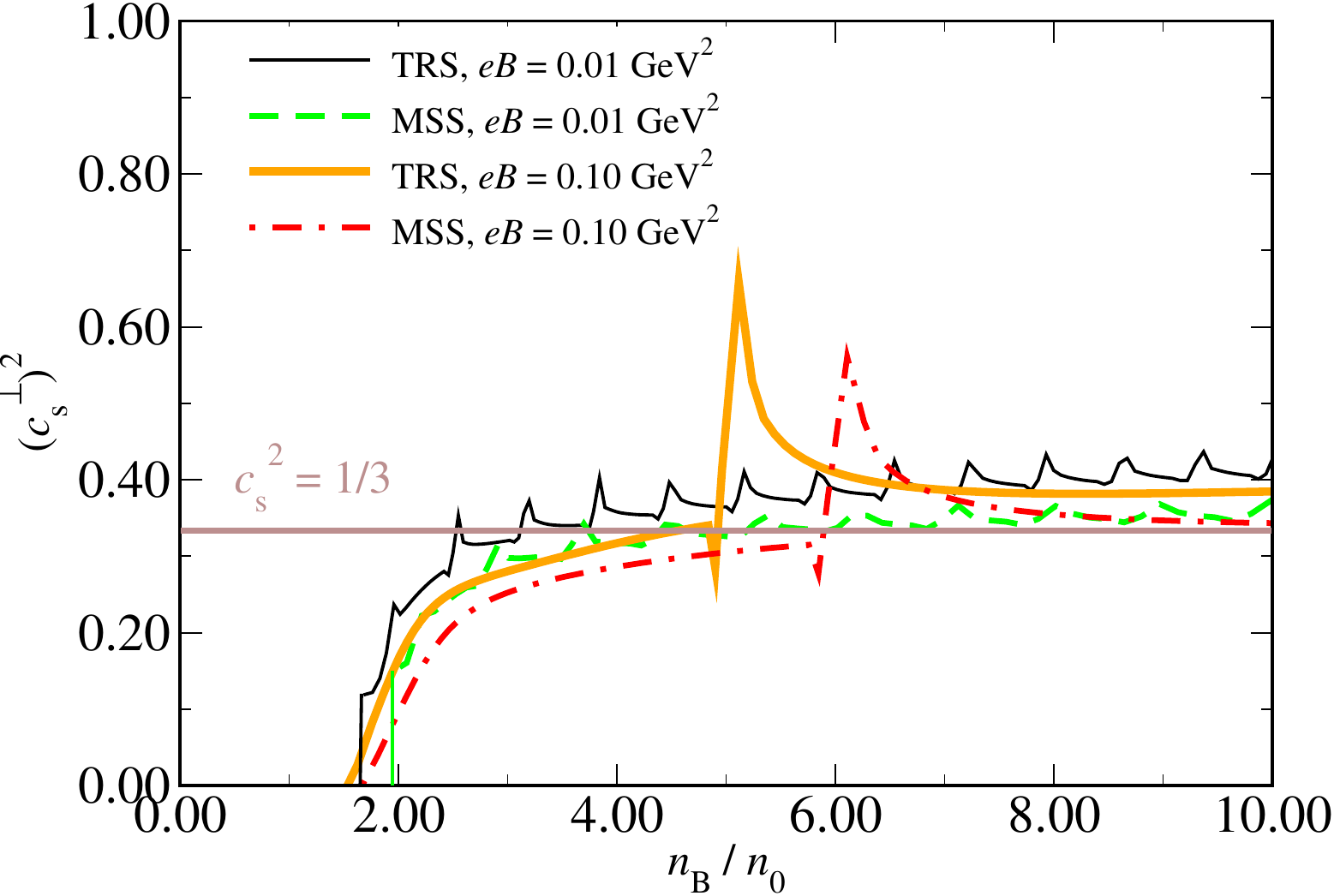}} 
 \caption{The components of squared sound velocity $ c_s^2 $ as a function of the normalized baryon number density $n_B / n_0 $, where $ n_0 $ denotes the nuclear saturation density. The figure shows the \textit{parallel} component $(c_s^\parallel)^2$ [panel (a)], while it shows the \textit{perpendicular} component  $(c_s^\perp)^2$ [panel (b)]. Results are shown for different regularization schemes in the presence of a finite magnetic field. The horizontal line indicates the limiting value $c_s^2 = 1/3$, characteristic of an ultrarelativistic gas.}
   \label{Fig7}
\end{figure}

Panels (a) and (b) of Fig.~\ref{Fig7} show  the parallel and perpendicular components of the squared speed of sound, $c_s^2$, as functions of the baryon number density. 
The quantization of Landau levels generates 
van Alphen-de Haas oscillations, producing discontinuities in the density of states. This effect is more pronounced at lower magnetic fields, where more levels contribute, leading to higher-frequency oscillations. For $eB = 0.10~\text{GeV}^2$, fewer levels contribute, and the behavior becomes smoother. Within MFIR, both TRS and MSS suppress spurious oscillations from regularization, while MSS preserves medium contributions near the Fermi surface, yielding physically consistent oscillations and a behavior for $ c_s^2 $ toward the conformal limit $1/3$. Furthermore, the qualitative dependence of the squared speed of sound on the baryon density agrees with recent nonlocal
Nambu--Jona-Lasinio (nlNJL) studies of cold magnetized quark matter~\cite{Ferraris:2025fva} and is also consistent with analyses that predict anisotropic sound velocities and a nonmonotonic behavior approaching the conformal limit in strongly magnetized dense matter~\cite{Ferrer:2022afu}.

\begin{figure}[htpb!]
\includegraphics[scale=0.33]{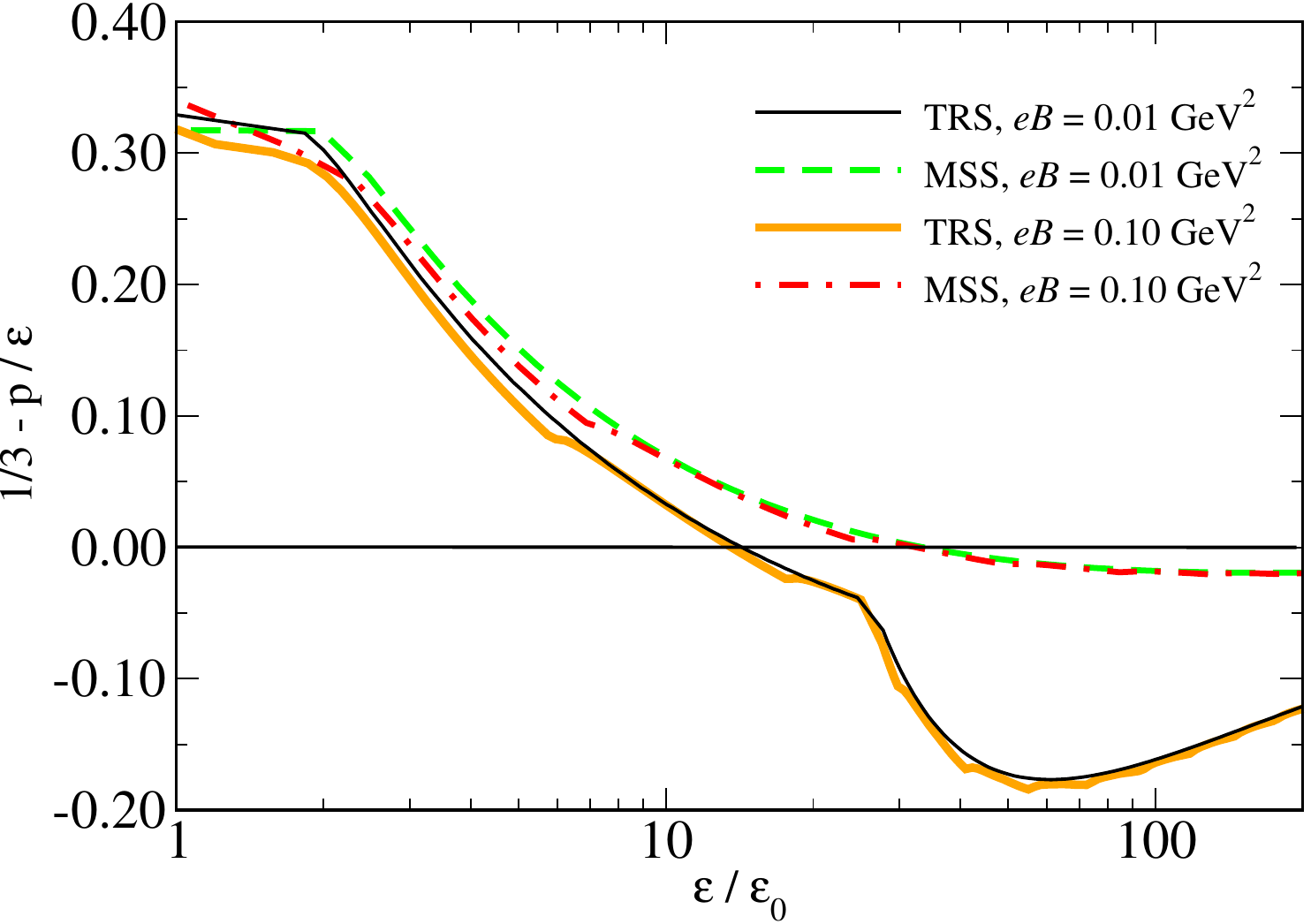}
 \caption{The trace anomaly as a function of the energy density $\varepsilon$ normalized by the energy density at nuclear saturation density $\varepsilon_0$ for a finite magnetic field.}
   \label{Fig8}
\end{figure}

An important thermodynamic quantity for probing conformality is the trace anomaly, defined as $ \mathfrak{T} = \varepsilon - 3p $, or, equivalently,
\begin{equation}
\frac{\mathfrak{T}}{3\varepsilon_N} = \frac{1}{3} - \frac{p_N^{\parallel}}{\varepsilon_N}.
\label{traceAnom}
\end{equation}
In QCD, this relation has contributions from both the vacuum and the medium. At zero temperature and finite baryon density, the medium contribution is particularly relevant and can be obtained directly from the equation of state. Following Ref.~\cite{Fujimoto:2022ohj}, the normalized form in~\eqref{traceAnom} can be used to quantify the deviations from conformality. In the magnetized NJL model, as may be seen in Fig.~\ref{Fig8}, MSS scheme produces a smoother and more rapid approach to the conformal limit $ \mathfrak{T} \to 0 $ at high densities when compared to the TRS. MSS approach not only eliminates regularization artifacts but also yields a monotonic decrease in the normalized trace anomaly with increasing energy density, consistent with recent neutron star constraints. This indicates that strongly interacting matter can approach conformality in the high-density regime, even under strong magnetic fields.
 
\section{Discussion and Final Remarks}
\label{Sec:remarks}

In this study, we have investigated the effects of different regularization schemes within the cold and dense NJL model in the presence of an external magnetic field. By systematically applying the MFIR alongside MSS, we ensured the proper regularization of vacuum contributions while keeping medium-dependent integrals unrestricted. This combined treatment guarantees a physically consistent separation between vacuum and medium effects, avoiding the introduction of spurious divergences and preserving the correct behavior of relevant degrees of freedom near the Fermi surface.

In the context of color superconducting phases under external magnetic fields, the application of MFIR is particularly important. The strong spurious oscillations observed in the order parameters when employing smooth functions dependent on $eB$ are frequently misinterpreted as the van Alphen-de Haas oscillations. While these genuine oscillations, associated with the filling of Landau levels at finite density, are present in the system, they are significantly less pronounced, as shown on solid and dashed curves of Fig.~\ref{Fig2}. The results obtained here clearly show that only the combination of MFIR and MSS fully removes the unphysical oscillations, leading to smooth and physically meaningful dependencies on both the magnetic field and density.

While MFIR alone successfully removes unphysical magnetic oscillations, it does not reproduce the correct asymptotic behavior of the diquark condensate $\Delta$ at high densities. The MSS implementation, on the other hand, ensures a monotonically increasing condensate with chemical potential, consistent with lattice QCD (for $N_c=2$), chiral perturbation theory, and renormalization-group analyses. The implementation of MSS at $eB=0$ as well as the combined application of MSS and MFIR therefore represents a robust framework to describe color superconductivity in magnetized and dense quark matter.

The explicit separation between vacuum and medium terms alters the phase structure of the system, shifting the critical lines in the $\mu \times eB$ plane and removing the artificial suppression of the superconducting phase at large chemical potentials. This improvement demonstrates the fundamental importance of a proper vacuum regularization when studying QCD-like models at high density.

The findings of this study reinforce the necessity of carefully handling regularization in NJL-like models, particularly in the presence of external fields and finite density effects. From a phenomenological perspective, the MSS framework paves the way for more reliable applications of the NJL model to astrophysical environments such as magnetars and neutron star mergers, where strong magnetic fields and color superconductivity coexist. Extensions to the finite temperature regime are currently being pursued, and they will be reported in subsequent works. Furthermore, comparative analysis with lattice QCD data and other nonperturbative approaches could provide deeper insights into the reliability and limitations of this scheme in describing the QCD phase diagram. In summary, the MSS + MFIR scheme emerges as a consistent and physically grounded regularization framework, capable of describing dense and magnetized quark matter without unphysical artifacts and in close agreement with first-principles expectations.

\begin{acknowledgments}
This work was partially supported by Conselho Nacional de Desenvolvimento Cient\'ifico e Tecno\-l\'o\-gico  (CNPq), Grants No. 312032/2023-4, No. 402963/2024-5 and 445182/2024-5 (R.L.S.F.); Funda\c{c}\~ao de Amparo \`a Pesquisa do Estado do Rio 
Grande do Sul (FAPERGS), Grants No. 24/2551-0001285-0 (R.L.S.F.), No. 23/2551-0000791-6 and No. 23/2551-0001591-9 (D.C.D.); Coordena\c{c}\~ao de
Aperfei\c{c}oamento de Pessoal de N\'ivel Superior - Brasil (CAPES) -
Finance Code 001 (F.X.A.). The work is also part of the project
Instituto Nacional de Ci\^encia e Tecnologia - F\'isica Nuclear e
Aplica\c{c}\~oes (INCT - FNA), Grants No. 464898/2014-5 and No. 408419/2024-5, and supported
by the Ser\-ra\-pi\-lhei\-ra Institute (Grant No. Serra -
2211-42230). R. L. S. F. acknowledges
the kind hospitality of the Center for Nuclear Research at Kent State University, where part of this work was
done. 
\end{acknowledgments}

\section*{ Data Availability} 

The data that support the findings of this article are not publicly available. The data are available from the authors upon reasonable request.

\bibliography{ref}

\end{document}